\documentclass[11pt,a4paper,english]{article}
\usepackage{jmlr2e}
\usepackage{geometry}                
\usepackage{color} 
\usepackage{amsmath}
\usepackage{amsfonts}
\usepackage{mathrsfs}
\usepackage{epstopdf}
\usepackage{array}

\usepackage{multirow}

\usepackage[lofdepth,lotdepth]{subfig}
\usepackage{tikz}
\usetikzlibrary{arrows,shapes}

\makeatletter
\def\captionof#1#2{{\def\@captype{#1}#2}}
\makeatother

 {\begin{quote}\color{red}\begin{sffamily}}%
 {\end{sffamily}\color{black}\end{quote}}

\newcommand{\bA}{{\bf A}}
\newcommand{\ba}{{\bf a}}
\newcommand{\bc}{{\bf c}}
\newcommand{\bC}{{\bf C}}
\newcommand{\bz}{{\bf z}}
\newcommand{\bv}{{\bf v}}
\newcommand{\bu}{{\bf u}}
\newcommand{\balpha}{{\boldsymbol \alpha}}
\newcommand{\bbeta}{{\boldsymbol \beta}}
\newcommand{\bdelta}{{\boldsymbol \delta}}
\newcommand{\bDelta}{{\boldsymbol \Delta}}
\newcommand{\bgamma}{{\boldsymbol \gamma}}

\newcommand{\beps}{{\boldsymbol \varepsilon}}

\newcommand{\bgammak}{\bgamma_{k}}
\newcommand{\bdeltak}{\bdelta_{k}}
\newcommand{\bbetak}{\bbeta_{k}}
\newcommand{\bbetaun}{{\bbeta_{\ell}}}

\newcommand{\betabar}{\overline {\beta}}
\newcommand{\bbetabar}{\overline {\boldsymbol \beta}}

\newcommand{\hbgammak}{{\widehat{\bgamma}_{k}}}

\newcommand{\hbbetak}{{\widehat{\bbeta}_{k}}}
\newcommand{\hbbeta}{\widehat {\boldsymbol \beta}}
\newcommand{\hbeta}{\widehat {\beta}}

\newcommand{\hbetabar}{\widehat {\betabar}}
\newcommand{\hbbetabar}{\widehat {\bbetabar}}
\newcommand{\hbgamma}{\widehat {\boldsymbol \gamma}}
\newcommand{\hbdelta}{\widehat {\boldsymbol \delta}}

\newcommand{\btheta}{{\boldsymbol \theta}}

\newcommand{\bomega}{{\boldsymbol \omega}}

\newcommand{\zz}{(\bz_k)_{k=1}^K}
\newcommand{\dv}{(\bv_k)_{k=1}^K}
\newcommand{\du}{(\bu_k)_{k=1}^K}

\newcommand{\dgamman}{{(\widehat{\bgamma}_k)_{k=1}^K}}
\newcommand{\dgamma}{{(\bgamma_k)_{k=1}^K}}
\newcommand{\dbeta}{({{\bbeta_k}})_{k=1}^K}
\newcommand{\dalpha}{({{\balpha_k}})_{k=1}^K}

\newcommand{\bB}{{\bf B}}
\newcommand{\bI}{{\bf I}}
\newcommand{\by}{{\bf y}}
\newcommand{\bY}{{\bf y}}
\newcommand{\bX}{{\bf X}}
\newcommand{\bx}{{\bf x}}

\newcommand{\LL}{{\mathcal L}}

\newcommand{\bcX}{{\boldsymbol {\mathcal X}}}

\newcommand{\bcY}{{\boldsymbol {\mathscr{Y}}}}

\renewcommand{\P}{{\rm I}\kern-0.12em{\rm P}}
\newcommand{\1}{{\rm 1}\kern-0.28em{\rm I}}
\newcommand{\E}{{\rm I}\kern-0.12em{\rm E}}
\newcommand{\R}{{\rm I}\kern-0.14em{\rm R}}
\newcommand{\N}{{\rm I}\kern-0.14em{\rm N}}
\newcommand{\reals}{{\rm I}\kern-0.16em{\rm R}}

\newcommand{\p}{{\rm I}\kern-0.18em{\rm P}}

\DeclareMathOperator*{\argmin}{argmin}

\tikzstyle{vertex1}=[circle,fill=blue, draw,minimum size=7pt,inner sep=0pt]
\tikzstyle{vertex2}=[circle,fill=black, draw,minimum size=7pt,inner sep=0pt]
\tikzstyle{vertex3}=[circle,fill=white, draw, minimum size=7pt,inner sep=0pt]
\tikzstyle{vertex4}=[circle,fill=red, draw, minimum size=7pt,inner sep=0pt]

\tikzstyle{edge} = [draw,thick,-]
\tikzstyle{weight} = [font=\small]

\title{Joint estimation of $K$ related regression models with simple $L_1$-norm penalties}
\author{
\name{ E. Ollier$^{\star}$ and V. Viallon$^\ddag$} \\ \\
\addr{
$^\star$ Universit\'e de Lyon, F-69622, Lyon, France \\
$^\ddag$Universit\'e de Lyon, F-69622, Lyon, France;
Universit\'e Lyon 1, UMRESTTE, F-69373 Lyon;\\
IFSTTAR, UMRESTTE, F-69675 Bron.\\
}
\email{vivian.viallon@univ-lyon1.fr}
\email{ed.ollier@gmail.com}
}

\begin{document}

\maketitle

\begin{abstract}
We propose a new approach, along with refinements, based on $L_1$ penalties and aimed at jointly estimating several related regression models. Its main interest is that it can be rewritten as a weighted lasso on a simple transformation of the original data set. In particular, it does not need new dedicated algorithms and is ready to implement under a variety of regression models, {\em e.g.}, using standard R packages. Moreover, asymptotic oracle properties are derived along with preliminary non-asymptotic results, suggesting good theoretical properties. Our approach is further compared with state-of-the-art competitors under various settings on synthetic data: these empirical results confirm that our approach performs at least similarly to its competitors. As a final illustration, an analysis of road safety data is provided.
\end{abstract}

\section{Introduction}

With the emergence of high-dimensional data, penalized regression models have now become standard, with such penalties as the $L_q$-norm or quasinorm of the parameter vector, for some $q\geq 0$. Underlying the use of these penalties, a very common assumption when working with moderate to high dimensional data is that the theoretical parameter vector $\bbeta^*$ is {\em sparse}: only a small subset of its $p$ components is expected to be non-null. Sparsity-inducing penalties, such as those relying on the $L_1$-norm, are especially useful in this context: the $L_1$-norm being convex it further leads to approaches that can generally be solved efficiently; they are often referred to as the lasso. Under appropriate conditions, lasso estimates especially attain optimal convergence rates (up to logarithmic terms): sparsity-inducing approaches are not only appealing for interpretation matters, but also because they can improve upon non-penalized procedures and lead to optimal estimates regarding estimation and/or prediction accuracy (see {\em e.g.}, \cite{BRT09, BvdG, Dalalyan14} for the linear regression case).

In many applications, the objective is actually to estimate several, $K$ say, parameter vectors $\bbeta^*_1,\ldots, \bbeta^*_K$ corresponding to $K$ related regression models; this problem is often referred to as multi-task learning in the literature \citep{AEP, MaurerPontil, NegahbanWainwright, LouniciTsyb, JacobVertBach}. Our motivating  example corresponds to a standard situation in epidemiology or clinical research, where data come from several {\em strata} of the population \citep{GertheissTutz, Viallon}. Each stratum can correspond to a type or dosage of treatment, a given geographical area or can be defined by crossing age category and gender, etc. The relationship between the response variable and the covariates has then to be studied on each stratum. This relationship is rarely strictly the same over all the strata, but some homogeneity is generally expected.  As a result, simple strategies consisting in estimating one model per stratum or one model on all the strata pooled together are generally sub-optimal in this context, and appropriate approaches should automatically adapt for the level of heterogeneity over the strata. Besides the sparsity assumption on each $\bbeta^*_k$,  vectors $\bbeta^*_k$ are usually expected to be close to each other in some sense. The principle of multi-task learning strategies is then to account for the presumed similarity among the $\bbeta^*_k$'s, while not masking potential heterogeneities, in order to improve estimation performance. Recent methods rely on penalties  encouraging estimators of matrix $\bB^* = (\bbeta^*_1,\ldots, \bbeta^*_K)\in\R^{p\times K}$ to exhibit some given structure: low-rank with the trace-norm penalty, group-structure with $L_1/L_2$ or $L_1/L_\infty$ penalties, etc.  To solve the corresponding optimization problem, dedicated algorithms are generally needed which limits their use and generalization by practitioners: most of the available algorithms enable to consider only the $L_2$ loss (for linear regression typically), and/or the logistic or hinge loss.  
In this paper we propose a new approach which is very intuitive and easy to implement under a variety of models, {\em i.e.}, for a variety of loss functions. Indeed, our approach reduces to a weighted lasso on a straightforward transformation of the data. As a result, any function or package enabling the resolution of the weighted lasso can directly be used for the implementation. In particular, the glmnet R package of \cite{glmnet} allows the treatment of linear, logistic, Poisson, multinomial and Cox models.  The principle of our approach is very simple: we encourage sparsity within individual $\bbeta^*_k$ and similarity between the $\bbeta^*_k$'s. As will be shown below, we reach this objective by simply using $L_1$-norm penalties and our approach returns estimates for matrix $\bB^*$ that are sums of a rank one matrix and a sparse one. 

In the following Section \ref{sec:Methods}, we formally introduce the setting we consider, describe our approach, state its connections with previous works, discuss its practical implementation and present theoretical results.   Then, results from simulation studies are presented in Section \ref{sec:Simulations},  where comparisons are made with state-of-the-art methods. Finally, an illustration of our approach is provided on road safety data in Section \ref{sec:RealData}, where $K=18$ experts have been asked to determine drivers' responsibility in road traffic accidents and the objective is to assess expert agreement. 


\section{Methods}\label{sec:Methods}

Although the main applications we have in mind concern data gathered from various strata of a population, our work falls into the more general multi-task learning context. For the sake of simplicity, we will use the terminology stratum everywhere, but everything can be extended by replacing stratum by task. In addition, methods as well as their theoretical properties will be presented in the linear regression model with no intercept for ease of notation. Extensions to generalized linear models \citep{McC89} are straightforward. How to deal with intercept terms in practice is briefly discussed in Section \ref{sec:Implementation}  below.

\subsection{Notations} For any integer $m\geq 1$, we will denote by $[m]$ the set of values $\{1,\ldots,m\}$. Moreover, ${\bf 0}_m$ and ${\bf 1}_m$ will denote the vectors of size $m$ with components all equal to 0 and 1 respectively, while ${\bf I}_m$ will stand for the $(m\times m)$ identity matrix. For any vector $\bx=(x_1,\ldots,x_m)\in\R^m$, we further denote by $supp(\bx)$ its support ({\em i.e.}, $supp(\bx) = \{j\in[m]: x_j \neq 0\}$) and we set $\|\bx\|_q = (\sum_{j\in[m]} |x_j|^q)^{1/q}$ for any real number $q\in(0,\infty)$, $\|\bx\|_\infty = \max_j |x_j|$ and $\|\bx\|_0 = |supp(\bx)|$, where $|E|$ denotes the cardinality of the set $E$. For any set $E\subseteq [m]$, we will denote by $\bx_E$ the vector of $\R^{|E|}$ with components $(x_j)_{j\in E}$. 

\noindent Denote by $K\geq 1$ the number of strata and by $n_k$ the number of observations in stratum $k\in[K]$, with $n=\sum_{k\in[K]}n_k$ the total number of observations. We will assume that for any $k\in[K]$ response vectors  $\by^{(k)} = (y^{(k)}_1,\ldots,y^{(k)}_{n_k})^T\in\R^{n_k}$ are related to design matrices $\bX^{(k)} = ({\bx^{(k)}_1}^T,\ldots,{\bx^{(k)}_{n_k}}^T)^T \in \R^{{n_k}\times p}$ according to the following linear model
\begin{equation}  
\by^{(k)} = \bX^{(k)} {\bbeta^*_k} + \beps^{(k)}.\label{eq:model}
\end{equation}
In this equation, vectors $\beps^{(k)} = (\varepsilon^{(k)}_1,\ldots,\varepsilon^{(k)}_{n_k})^T\in\R^{n_k}$ denote noise vectors:  variables $\varepsilon^{(k)}_i$ will be assumed to be independent and identically distributed (i.i.d.) with $\E \varepsilon^{(k)}_i = 0 $ and Var$(\varepsilon^{(k)}_i) = \sigma^2$ for some unknown $\sigma^2>0$, for all $i\in[n_k]$ and $k\in[K]$. Finally, vectors $\bbeta^*_k\in\R^p$ are the $K$ parameter vectors of interest, to be estimated. 
In the context considered here, the $\bbetak^*$'s are expected to be sparse and close to each other in some sense.

\subsection{Proposed approach}\label{Sec:ourproposal}

\subsubsection{Principle}
Our proposal relies on the following decomposition for $\bbetak^*$, $k\in[K]$:
\begin{equation} 
\bbetak^* = \bbetabar^* + \bgammak^*.  \label{eq:decomp_add}
\end{equation}
Here $\bbetabar^*$ describes what is ``common'' over the strata, while $\bgammak^*$ captures the variation in stratum $k$ around $\bbetabar^*$. Of course, there are infinitely many such decompositions, but assuming that vectors $\bbeta_k^*$ are close to each other, those minimizing  the quantity $\sum_k \|\bgammak^*\|_q$, for some $q\geq 0$, are naturally appealing. They correspond to very natural choices for vector $\bbetabar^* \in {\rm arg}\min_{\bbetabar} \sum_k \|\bbetak^* - \bbetabar\|_q$. In particular, it is straightforward that choices $q=0, 1$ and $2$ correspond to decompositions where, for each $j\in[p]$, $\betabar^*_j$ is a mode, a median and the mean of $(\beta^*_{1,j},\ldots, \beta^*_{K,j})$, respectively. 

The principle of our proposal is to obtain estimators of the form $\hbbetak= \hbbetabar+ \hbgammak$ with sparse $\hbbetabar$ and sparse $\hbgammak$. For appropriate values $\lambda_1\geq 0$ and $\lambda_{2,k}\geq 0$, our estimates are then defined as 
\begin{align}
\hspace{-30pt}(\hbbetabar,\hbgamma_1\ldots,\hbgamma_K)\in \argmin_{\bbetabar,\bgamma_1,\ldots,\bgamma_K}\left\{\sum_{k\geq 1} \frac{\|\bY^{(k)} - \bX^{(k)}(\bbetabar+\bgamma_{k})\|_2^2 }{2n}
  + \lambda_1 \|\bbetabar\|_1 + \sum_{k\geq 1} \lambda_{2,k}\|\bgamma_{k}\|_1\right\}.\label{est_M1}
\end{align}
Clearly, $\hbbetabar={\bf 0}$ and $\hbgammak=\hbbetak$ for $\lambda_1$ large enough:
the $K$ problems are solved independently (the corresponding strategy will be referred to as IndepLasso in the sequel). On the other hand, we have $\hbbetak =\hbbetabar$ for all $k$ if the $\lambda_{2,k}$'s are large enough: optimal solutions are those of the lasso run on all the data pooled together (the corresponding strategy will be referred to as IdentLasso in the sequel). This is a nice property since selecting appropriate values for $\lambda_1$ and the $\lambda_{2,k}$'s allows us to adapt to the extent of homogeneity over the strata. We refer to  Section \ref{sec:TuningParamSel} below for a discussion on how to chose $\lambda_1$ and $\lambda_{2,k}$ in practice. 

\noindent The optimization problem described in Equation (\ref{est_M1}) is of course equivalent to the minimization of 
\begin{align}
\sum_{k\geq 1} \frac{\|\bY^{(k)} - \bX^{(k)}\bbetak\|_2^2 }{2n}
  +  \lambda_1 \{\|\bbetabar\|_1 + \sum_{k\geq 1} \frac{\lambda_{2,k}}{\lambda_1}\|\bbetak - \bbetabar\|_1\} \label{eq:RewritingM1}
\end{align}
over $\bbetak\in\R^p$ for $k\in[K]$ and $\bbetabar\in\R^p$. Because $\bbetabar$ now only appears in the penalty terms, it is easy to see that, at optimum, $\hbetabar_j$ is a weighted and shrunk version of the median of $(\hbeta_{1,j},\ldots, \hbeta_{K,j})$; we will denote it by ${\rm WSmedian}_{\mu_{[K]}}(\hbeta_{1,j},\ldots, \hbeta_{K,j})$ with $\mu_{[K]} = (\mu_1,\ldots,\mu_K)$ and $\mu_k = \lambda_{2,k}/\lambda_1$. If all $\mu_k$'s are equal and tend to 0, then ${\rm WSmedian}_{\mu_{[K]}}$ tends to the constant function returning ${\bf 0}_p$: this corresponds to the decomposition $\hbbetak = \hbgamma_k$, and then to the IndepLasso strategy. On the other hand,  if all $\mu_k$'s are equal but tend to infinity, then ${\rm WSmedian}_{\mu_{[K]}}$ tends to the standard median function: this corresponds to the decomposition $\hbbetak = \hbbetabar + \hbgamma_k$, where $\hbetabar_j = {\rm median}(\hbeta_{1,j}, \ldots, \hbeta_{K,j})$. If the $\mu_k$'s are all equal to $1/\tau$, with $\tau \in \N$, then ${\rm WSmedian}_{\mu_{[K]}}$ is a shrunk median in the sense that
\begin{align*}
\hbetabar_j 
= {\rm WSmedian}_{1/\tau,\ldots,1/\tau}(\hbeta_{1,j},\ldots, \hbeta_{K,j})
= {\rm median}({\bf 0}_\tau^T, \hbeta_{1,j},\ldots, \hbeta_{K,j}).
\end{align*}
A last interesting example is when for some $\ell\in[K]$ $\mu_\ell\rightarrow \infty$ and the other $\mu_k$'s are fixed (finite), for all $k\neq \ell$. Then ${\rm WSmedian}_{\mu_{[K]}}(\hbeta_{1,j},\ldots, \hbeta_{K,j})\rightarrow \hbeta_{\ell,j}$, leading to the decomposition $\hbbetak = \hbbeta_\ell + \hbgamma_k$, with $\hbgamma_\ell = {\bf 0}_p$: this corresponds to considering stratum $\ell$ as the reference one (see Section \ref{sec:adapM1} below). To sum up, each particular choice for the $\lambda_{2,k}/\lambda_1$ ratios ``identifies'' a particular common effect vector $\hbbetabar$ with $\hbetabar_j$ defined as ${\rm WSmedian}_{\mu_{[K]}}(\hbeta_{1,j},\ldots, \hbeta_{K,j})$, and our approach encourages solutions $(\hbbeta_{1},\ldots, \hbbeta_{K})$ with typically sparse vector of common effects $\hbbetabar$ and sparse vectors of differences $\hbbetak - \hbbetabar$.

\subsubsection{Implementation: rewriting as a weighted lasso}\label{sec:Implementation}
A very attractive property of our approach is that it can be rewritten as a simple weighted lasso. To state this, first set $ \bcY = ({\bY^{(1)}}^T,\ldots,{\bY^{(k)}}^T)^T\in\R^n$ and 
introduce the quantities
$$  
\bcX = \left( 
\begin{array}{c c c c }
\bX^{(1)} &\bX^{(1)}& \hdots &{\bf 0}  \\
\vdots & \vdots & \ddots & \vdots \\
\bX^{(K)} &{\bf 0}&\hdots  &  \bX^{(K)}  
\end{array}
\right) \quad{\rm and} \quad
\btheta = \left( 
\begin{array}{c}
\bbetabar\\
\bgamma^{(1)}\\
\vdots\\
\bgamma^{(K)}
\end{array}
\right),
$$ which are elements of $\R^{n\times (K+1)p}$ and $\R^{(K+1)p}$ respectively.
Further introduce the vector of weights $\bomega = ({\bf 1}_p^T,(\lambda_{2,1}/\lambda_1){\bf 1}_p^T,\ldots,(\lambda_{2,K}/\lambda_1){\bf 1}_p^T)^T\in\R^{(K+1)p}$.  Then, setting $\|{\btheta}\|_{1,\bomega} = \sum_{j=1}^{(K+1)p} \omega_j |\theta_j|$, the criterion to be minimized in Equation (\ref{est_M1}) can be rewritten as 
\begin{equation}
 \frac{ \|\bcY - \bcX\btheta\|_2^2}{2n} + \lambda_1\|{\btheta}\|_{1,\bomega} \label{eq:M1_Lasso}
\end{equation}
which is the criterion minimized in the weighted lasso. Of course, it is easy to show that this remains true under generalized linear models, Cox models, etc. As mentioned above, this is particularly interesting since it means that our approach can be implemented under a variety of models using available packages for the lasso, {\em e.g.}, the glmnet R package of \cite{glmnet}.

Throughout the paper, only linear regression models with no intercept are considered for ease of notation. In practice however, intercept has generally to be included in the model. A first option consists in using the glmnet function with ``intercept=FALSE'' and replace ${\bf X}^{(k)}$ by ${\bf Z}^{(k)} = ({\bf 1}_{n_k},{\bf X}^{(k)})\in\R^{n_k\times(p+1)}$ in the definition of $\bcX$ above: this way, the absolute value of the common intercept term is penalized, and so are variations around this common intercept. In many situations, it makes more sense not to penalize the common intercept: an alternative option is then to use the glmnet function with ``intercept=TRUE'' and replace ${\bf X}^{(k)}$ by ${\bf Z}^{(k)}$ in the definition of $\bcX$ everywhere except in the first $p$ columns of $\bcX$: this way, only variations around the common intercept (which corresponds to the median of the intercepts over the strata in this case) are penalized.

\subsubsection{Related work}\label{sec:RelatedWork} 


Decomposition (\ref{eq:decomp_add}) was first suggested in \cite{EvgeniouPontil} who use the SVM machinery with $L_2$-norm penalties. For the application we have in mind, using sparsity-inducing norms as the $L_1$-norm is more appealing. Indeed, detecting the subset of covariates that have a differential effect over the strata is often of primary interest (see our application in Section \ref{sec:RealData} below). By using the $L_1$-norm of the $\bgamma_k$'s our approach enjoys good properties regarding support recovery for both the $\bbeta_k$'s and $\bgamma_k$'s (as will be shown below), which is of course not the case for methods based on $L_2$-norm penalties. Another advantage when using $L_1$-norms is that the common effect estimation is more robust (median versus mean). 

\cite{Jalali} consider a closely related decomposition, this time for the parameter matrix $\bB = (\bbeta_1,\ldots,\bbeta_K)\in\R^{p\times K}$. They write ${\bB} = {\bf R} + {\bf S}$, where ${\bf R}$ and ${\bf S}$ are two $(p\times K)$ matrices with element $(j,k)$ denoted by $r^{(k)}_j$ and $s^{(k)}_j$, respectively. Further denote by ${\bf r}^{(k)}$ and ${\bf s}^{(k)}$ the $k$-th column of matrices ${\bf R}$ and ${\bf S}$ respectively, so that $\bbeta^{(k)} = {\bf s}^{(k)} + {\bf r}^{(k)}$, and by ${\bf r}_{j}$ and ${\bf s}_{j}$ the $j$-th row of matrices ${\bf R}$ and ${\bf S}$. Further set, for any matrix ${\bf M}\in\R^{p\times K}=({\bf m}_1,\ldots,{\bf m}_p)^T$ and $q\geq 1$, $\|{\bf M}\|_{1,q} = \sum_{j\in[p]} \|{\bf m}_j\|_q$. Then their approach returns an estimate for $\bB^*$ derived from minimizers of the following criterion: 
\begin{equation} \sum_k \frac{ \|\bY^{(k)} - \bX^{(k)}({\bf r}^{(k)} + {\bf s}^{(k)})\|_2^2 }{2n} + \lambda_s \| {\bf S}\|_{1,1}+\lambda_r \| {\bf R}\|_{1,\infty}.\label{eq:dirty}
\end{equation}
The term $ \| {\bf S}\|_{1,1}$ encourages matrix ${\bf S}$ to be sparse (few elements are non-zero) while $ \| {\bf R}\|_{1,\infty}$ encourages matrix ${\bf R}$ to have a row-wise group structure (few lines have zero entries): together, they encourage solutions $\widehat\bB$ that can be written as the sum of a row-sparse matrix and a sparse one. As mentioned in \cite{Jalali}, setting $d = \lfloor \lambda_r/\lambda_s\rfloor$ , the combination of the two penalties leads to solutions $\widehat {\bf R}$ such that for any $j$ with $\|\widehat {\bf r}_j\|_\infty >0$ we have $|M_j| \geq d+1$ where $M_j = \{k: |\widehat r_j^{(k)}| = \|\widehat {\bf r}_j\|_\infty\}$ (see their Lemma 2): in other words, rows that are not uniformly null have at least $d+1$ components that have equal absolute values. In contrast, our approach returns estimates for $\bB^*$ that is a sum of a rank one matrix (each row has $p$ equal components instead of at least $d+1$ components of equal absolute values) plus a sparse matrix: it is therefore less flexible, but easier to interpret and, above all, to implement. From a theoretical point of view, Jalali et al. especially considered the case where $K=2$ and covariates are generated from a standard multivariate Gaussian distribution. Further assuming that $n_1=n_2 $, they show that the number of samples needed to ensure support recovery with high probability ({\em i.e.}, the sample complexity) is inferior for Dirty, compared to both IndepLasso (where $K$ lasso are run independently on the $K$ strata) and the group-lasso strategy relying on 
the $L_1/L_\infty$ penalty \citep{NegahbanWainwright}. As shown Section \ref{sec:SampleComp} of the Appendix, the sample complexity of our approach is never superior to that of Dirty if in addition to the assumptions considered in Jalali et al., $\beta^*_{1,j}\beta^*_{2,j}\geq 0$ for all $j\in[p]$. Our empirical results presented in Section \ref{sec:Simulations} further suggest that it may still be the case for $K>2$. Stating this theoretically is out of the scope of the present paper and will be considered elsewhere.

\subsection{Adaptive version and other refinement}

\subsubsection{Adaptive version}\label{sec:adapM1}

An adaptive version of our approach can be derived by selecting appropriate weights and then replacing the $L_1$ norms by weighted $L_1$-norms in Equation (\ref{est_M1}). Following the ideas of the adaptive lasso \citep{Zou06}, these weights can be constructed from initial estimators of $\betabar_j$ and $\gamma_{k,j} = \beta_{k,j} - \betabar_j$, for $j\in[p]$ and $k\in[K]$. Denoting by $\widetilde\bbeta_{k}$ initial estimates of $\bbeta_{k}$ ({\em e.g.}, OLS estimates or MLEs if $n_k\gg p$ for all $k\in[K]$), and  defining for any $j$, $$\ell_j = \min\{ k: \widetilde\beta_{k,j}\in {\rm median}(\widetilde\beta_{1,j},\ldots,\widetilde\beta_{K,j})\},$$ initial estimators for $\betabar_j$ and  $\gamma_{k,j} $ are set to $\widetilde\beta_{\ell_j,j}$ and $\widetilde\beta_{k,j}-\widetilde\beta_{\ell_j,j}$.
Note that because values $\widetilde\beta_{1,j},\ldots,\widetilde\beta_{K,j}$ are generally all distinct, their median can be a range of values if $K$ is even. In this case, we want to set the initial estimator of the common effect of covariate $j$ to one of the values in $(\widetilde\beta_{1,j},\ldots,\widetilde\beta_{K,j})$, hence the use of the minimum in the definition of $\ell_j$  (alternatively, the maximum could be used: there is no reason for favoring one or the other, at least a priori).
Then setting, for all $k\in[K]$ and $j\in[p]$, $w_{k,j} = 1/|\widetilde\beta_{k,j}|^\rho$ and $\nu_{k,j} = 1/|\widetilde\beta_{k,j}- \widetilde\beta_{\ell_j,j}|^\rho$, for some $\rho> 0$ (a typical value is $\rho=1$), the adaptive version of our approach consists in minimizing the following objective function
\begin{equation*}
\sum_{k\geq 1}\frac{\|\bY^{(k)} - \bX^{(k)}(\bbetabar + \bgammak)\|_2^2 }{2n} + 
\lambda_{1} \sum_{j=1}^p  w_{\ell_j,j} |\betabar_{j}|+ \sum_{k\geq 1}\lambda_{2,k}\sum_{j=1}^p \nu_{k,j} |\gamma_{k,j}|\Big\}. 
\end{equation*}
But because $\nu_{\ell_j,j}=+\infty$ for all $j\in[p]$, this is equivalent to minimizing the criterion: 
\begin{equation}
\sum_{k\geq 1}\frac{\|\bY^{(k)} - \bX^{(k)}\bbeta_k\|_2^2 }{2n} + 
\lambda_{1} \sum_{j=1}^p  w_{\ell_j,j} |\beta_{\ell_j,j}|+ \sum_{k\geq 1}\lambda_{2,k}\sum_{j=1}^p \nu_{k,j} |\beta_{k,j}-\beta_{\ell_j,j}|\Big\}. \label{est_adaM1}
\end{equation}

In other words, the adaptive version of our approach can be seen as a refinement of the following strategy, that is very common in epidemiology and clinical research when data come from several strata. Many practitioners would first select a reference stratum, say $\ell\in[K]$, and then use the decomposition: $\bbetak = \bbetaun + \bdeltak$, for any $k\neq \ell$. Then, a lasso can be used and parameter estimates for each stratum can be obtained from minimizers of the following objective function:
\begin{equation}
 \frac{1}{2n}\Big\{\|\bY^{(\ell)} - \bX^{(\ell)}\bbetaun\|_2^2 +  \sum_{k\neq\ell} \|\bY^{(k)} - \bX^{(k)}(\bbetaun+\bdelta^{(k)})\|_2^2 \Big\} + \lambda_1 \|\bbetaun\| + \sum_{k\neq \ell} \lambda_{2,k}\|\bdelta^{(k)}\|_1.   \label{eq:InterLasso}
 \end{equation}
This approach reduces to a standard lasso where to the original vector of covariates is augmented by interaction terms between the covariates and indicator functions describing membership to each stratum $k\neq \ell$. The adaptive version of our approach enjoys two advantages compared to this strategy, referred to as InterLasso hereafter. First, the selection of the reference stratum is automatic with our approach and based on initial estimates of the $\bbetak^*$,'s. Second, and above all, the reference stratum $\ell_j$ is covariate-specific. Whenever the initial estimates are consistent, $\ell_j$ belongs to the set ${\cal L}^*_j=\{ k: \beta^*_{k,j}\in {\rm median}(\beta^*_{1,j},\ldots, \beta^*_{K,j})\}$ with high probability for $n$ large enough (if the $n_k$'s all tend to $\infty$ as $n\rightarrow \infty$), and is therefore an appealing reference stratum for covariate $j$. 
As a result, our approach will generally yield models with lower complexity and hence better performance than the simple InterLasso (even if InterLasso can of course lead to better models than our approach in some situations; {\em e.g.}, if there exists a stratum $\ell$ for which $\beta^*_{\ell,j}\in {\rm mode}(\beta^*_{1,j}, \ldots,\beta^*_{K,j})$ for all $j\in[p]$, and this stratum $\ell$ is chosen as the reference one, and ${\rm mode}(\beta^*_{1,j}, \ldots,\beta^*_{K,j}) \cap {\rm median}(\beta^*_{1,j}, \ldots,\beta^*_{K,j})=\emptyset$ for some $j\in[p]$).

\subsubsection{Refinement}
With our first approach, sparsity of vectors $\hbbetak = \hbbetabar +\hbgammak$ is not directly encouraged, and is only ``induced'' by the sparsity of the common effects $\hbbetabar$ and of the variations $\hbgammak$. For instance, if for some $a\neq 0$ and $j\in[p]$, $\beta^*_{\ell_j,j} = a$ and $\beta^*_{k,j} = 0$ for some $k\neq \ell$, our first approach will typically not return $\widehat \beta_{k,j} = 0$ (unless it also returns $\widehat \beta_{\ell_j,j} = 0$).
We therefore propose a refined version which directly penalizes the $L_1$-norms of the $\bbetak$'s. 
More precisely, a second set of estimators can be defined as minimizers of the following criterion
\begin{equation}
\sum_{k\geq 1}\Big\{ \frac{\|\bY^{(k)} - \bX^{(k)}\bbetak\|_2^2 }{2n} + 
\lambda_{1,k} \|\bbetak\|_1+ \lambda_{2,k}\|\bbetak - \bbetabar\|_1\Big\} \label{est_M2}
\end{equation}
over $\bbetak\in\R^p$ for $k\in[K]$ and $\bbetabar\in\R^p$. In the sequel, this refined version will be referred to as $M_2$, while $M_1$ will refer to our first approach. Because  $\bbetabar$ now only appears in the last term of the objective function, it is clear that the $j$-th component of any optimal solution $\hbbetabar$ is such that 
$\hbetabar_j = {\rm median}(\hbeta_{1,j},\ldots, \hbeta_{K,j})$: for each covariate $j\in[p]$, its estimated common effect $\hbetabar_j$ corresponds to the median of its estimated effects over all the strata, irrespective to $\lambda_{1,k}/\lambda_{2,k}$ ratios. 


The implementation of $M_2$ is however less trivial than that of our first approach. In Section \ref{sec:ImplementationM2} of the Appendix, we derive the dual formulation of the optimization problem described in Equation (\ref{est_M2}), which is shown to reduce to a standard optimization problem for the linear regression (quadratic programming) and the logistic regression (entropy maximization problem) and can therefore be solved with available optimization toolboxes; the hinge loss can also be easily treated.  

An adaptive version of $M_2$ can further be recast in the generalized fused lasso framework \citep{Hol10, Viallon}. This makes its implementation straightforward with packages dedicated to the generalized fused lasso ({\em e.g.}, the FusedLasso R package of \cite{Hol10}). Recall the notations introduced in Section \ref{sec:adapM1} above. The adaptive version of $M_2$ consists in minimizing the following objective function
\begin{equation}
\sum_{k\geq 1}\Big\{ \frac{\|\bY^{(k)} - \bX^{(k)}\bbeta_k\|_2^2 }{2n} + 
\lambda_{1,k} \sum_{j=1}^p  w_{k,j} |\beta_{k,j}| + \lambda_{2,k}\sum_{j=1}^p \nu_{k,j} |\beta_{k,j}-\beta_{\ell_j,j}|\Big\} . \label{est_adaM2}
\end{equation}

Here again, we have $\nu_{\ell_j,j} = +\infty$ for all $j\in[p]$. To see the connection with the generalized fused lasso set $ \bcY = ({\bY^{(1)}}^T,\ldots,{\bY^{(k)}}^T)^T\in\R^n$ as above, and introduce $\bcX_F$ the $(n\times Kp)$ block diagonal matrix with $k$-th block equal to $\bX^{(k)}$. Further define ${\bf b} = (\bbeta_{1}^T,\ldots,\bbeta_{K}^T)^T = (b_1,\ldots,b_{Kp})\in\R^{pK}$. Then, the criterion of Equation (\ref{est_adaM2}) can be rewritten as 
$\|\bcY -\bcX_F{\bf b}\|_2^2/(2 n) + \lambda_1 \|{\bf b}\|_1 
+\lambda_2\sum_{j_1\sim j_2} |b_{j_1} - b_{j_2}|.$
Condition $j_1\sim j_2$ indicates that components $b_{j_1} $ and $b_{j_2} $ are connected in a particular graph that describes absolute differences that are penalized. Here, this graph consists of $p$ star-graphs where the $j$-th star-graph has coefficient $\beta_{\ell_j,j}$ at its center, that is connected to each $\beta_{k,j}$ for $k\neq \ell_j$. The adaptive version of $M_1$ can also be seen as a particular case of the generalized fused lasso, using the same graph made of $p$ star-graphs (our two adaptive versions penalize the same absolute differences), but where only the absolute value of the central node $|\beta_{\ell_j,j}|$ of each star-graph is penalized (while all the $|\beta_{k,j}|$'s are penalized in Equation (\ref{est_adaM2}) above). Finally note that the InterLasso strategy can also be seen as a special case of the generalized fused lasso: the underlying graph is made of $p$ star-graphs with $\beta_{\ell,j}$ as the central node (instead of $\beta_{\ell_j,j}$), and only the $|\beta_{\ell,j}|$ terms are penalized, for $j\in[p]$. 

These observations formally establish a connection with another strategy that was considered in \cite{GertheissTutz} and \cite{Viallon}. It will be referred to as CliqueFused, and it corresponds to a generalized fused lasso using this time a graph made of $p$ cliques (instead of $p$ star-graphs): for each covariate $j$, all absolute differences $|\beta_{k,j} - \beta_{k',j}|$ are penalized for all $k\neq k'$.  

The rewriting of the adaptive version of  $M_2$ as a special case of the generalized fused lasso shows that it is easy to implement with available packages, like the FusedLasso R packages for linear and logistic regression models; the same naturally holds for the adaptive version of $M_1$, but since it can still be written as a particular case of the (adaptive) lasso its implementation is faster using the glmnet function for instance. In addition asymptotic oracle properties for the adaptive version of  $M_2$ (and $M_1$) readily follow as direct consequences of the results obtained in \cite{Viallon} (see Section \ref{sec:Theo}  below); for the adaptive version of $M_1$, results obtained by \cite{Zou06} for the adaptive lasso can also be used.

\subsection{Theoretical results}\label{sec:Theo}
For the sake of brevity, we only present here a summary of asymptotic oracle properties for the adaptive versions of our approach. We refer to Section \ref{Theo_Supp} in the Appendix for more details along with preliminary non-asymptotic results for our first approach $M_1$.

We consider the situation where $p$ and $K$ are held fixed (they do not increase with $n$). We further assume that, for each $k\in[K]$, matrix ${\bX^{(k)}}^T \bX^{(k)}/n_k$ converges to a positive definite matrix $\bC^{(k)}$ as $n_k\rightarrow \infty$, and  $n_k/n\rightarrow \kappa_k$ as $n\rightarrow \infty$, with $0<\kappa_k<1$ (i.e., {\em stratum} sizes all tend to infinity at the same rate). These assumptions essentially imply that MLEs $\widetilde \beta_{k,j}$ exist and  are $\sqrt{n}$-consistent as $n\rightarrow\infty$; they will therefore be used in the definition of the weights in Equations (\ref{est_adaM1}) and (\ref{est_adaM2}). Our results essentially show that the adaptive version of $M_2$, say $M_2^{ad}$, enjoys {\em asymptotic oracle properties}. In particular, setting for any $j\in[p]$, $\ell_j^* = \min\{k: \beta^*_{k,j} \in{\rm median}(\beta^*_{1,j},\ldots,\beta^*_{K,j})\}$ and $K^*_{{\cal A}^*,j} =  \{k: \beta^*_{k,j}=\beta_{\ell^*_j,j}^*\}$ if $\beta_{\ell^*_j,j}^*\neq 0$, then all the parameters $\beta^*_{k,j}$ for $k\in K^*_{{\cal A}^*,j}$ are estimated by the common value $\hbeta^{(ad)}_{\ell_j,j}$, with probability tending to 1 as $n\rightarrow\infty$. This estimator has the same Gaussian limit distribution as that of the estimator we would obtain by pooling all data  corresponding to covariate $j$ and strata in $K^*_{{\cal A}^*,j}$ together. Moreover, $M_2^{ad}$ is asymptotically optimal in situations where, for all $j\in[p]$, $\beta^*_{k_1,j}=\beta^*_{k_2,j}$ implies that either $\beta^*_{k_1,j}=\beta^*_{k_2,j}=0$ or $\beta^*_{k_1,j}=\beta^*_{k_2,j}=\beta^*_{\ell_j^*,j}$, as in examples 1 to 3 of Figure \ref{fig:confbetas} in the Appendix. Asymptotic oracle properties can easily be derived for method $M_1^{ad}$ as well but, for $M_1^{ad}$ to be optimal we must have in addition: for all $j\in[p]$, $\{\beta^*_{\ell^*_j,j}\neq 0\} \Rightarrow \{\forall k\in[K], \beta^*_{k,j}\neq 0\}$. For instance, in example 2 on Figure \ref{fig:confbetas}, the estimator of $\beta^*_{1,j}$ returned by $M_1^{ad}$ can not be 0 (unless $M_1^{ad}$ returns a zero estimate for $\beta^*_{k_j,j}$ as well). Therefore, $M_1^{ad}$ would typically return a model with overall complexity higher than the theoretical one, hence be sub-optimal.  On the other hand, contrary to the CliqueFused strategy, both our approaches use star-graphs and are therefore sub-optimal in situations where non-zero values in $(\beta^*_{1,j},\ldots,\beta^*_{K,j})$ consist of at least two groups of identical values (and possibly some distinct non-zero values). For instance, in example 4 of Figure \ref{fig:confbetas}, neither $M_1^{ad}$ nor $M_2^{ad}$ can return non-null equal values for components $\hbeta^{(ad)}_{1,j}$, $\hbeta^{(ad)}_{2,j}$, and $\hbeta^{(ad)}_{3,j}$ (nor for $\hbeta^{(ad)}_{8,j}$, $\hbeta^{(ad)}_{9,j}$, and $\hbeta^{(ad)}_{10,j}$), while CliqueFused typically would, for $n$ large enough. However, CliqueFused may of course be outperformed by our approaches on finite samples. This was illustrated in the simulation study conducted in \cite{Viallon} in the single-task setting: they evaluated the robustness of the generalized fused lasso to graph misspecification and especially observed that the clique-based strategy, though asymptotically optimal, was outperformed by other graphs-based strategies on finite samples. In the present multi-task setting, if $\beta^*_{1,j},\ldots,\beta^*_{K-1,j}$ are all equal and $\beta^*_{K,j}$ is different from the $K-1$ other ones, we would have $\ell_j\in[K-1]$ for $n$ enough, and our approaches would only penalize $|\beta_{K,j} - \beta_{\ell_j,j}|$: they are therefore more likely to detect this difference than CliqueFused which penalizes all the differences $|\beta_{K,j} - \beta_{k,j}|$ for all $k\in[K-1]$.

\section{Simulation Study}\label{sec:Simulations}

\subsection{Competing Methods - Implementation}
Our main objective here is to illustrate the two new approaches introduced in Section \ref{sec:Methods} and compare them with state-of-the-art competitors, in particular with those presented above which show some links with our proposal: CliqueFused and Dirty. For comparison, we further included IndepLasso and IdentLasso. The sparse group-lasso strategy relying on the $L_1 + L_1/L_\infty$  penalty and referred to as spGroupLasso hereafter is considered too (see Section \ref{sec:Group} of the Appendix for a brief description of this approach and \cite{NegahbanWainwright} and \cite{SimonSpGpLasso} for more details). Results obtained with the stepwise method described in \cite{GertheissTutz} will also be presented in the low-dimensional example.
 
The glmnet R function of \cite{glmnet} was used to implement IndepLasso, IdentLasso as well as our approach $M_1$ (and its adaptive version). As for $M_2$, we used the Mosek Matlab toolbox (available at www.mosek.com) to solve the dual formulation presented in the Supplementaty Material. The FusedLasso R package of \cite{Hol10} was used to implement the CliqueFused approach and the adaptive version of $M_2$.  The spGroupLasso strategy was implemented with the spams R package of \cite{SPAMS}, and we used the gvcm.cat R package for the stepwise approach of \cite{GertheissTutz}. Finally, we used the script of \cite{Jalali} (available at http://ali-jalali.com/index\_files/L1Linf\_LASSO.r) to implement Dirty. However, this script actually implements a revised version of Dirty. At each iteration $\iota$ of their algorithm, tuning parameters are divided by $\sqrt{\iota}$ in the coordinate gradient descent steps, making the soft-thresholding operator closer and closer to the hard-thresholding one as $\iota$ increases. This trick then returns solutions that are barely shrunk and is related in some way to the relaxed lasso of \cite{M07}, the adaptive lasso of \cite{Zou06} and $L_q$-penalization with $0\leq q<1$. For this reason, solutions returned by this script will be referred to as RevDirty. Note that the trick of RevDirty could of course be used for our approach if a lasso function existed with this trick; we do not know any such function or package though. RevDirty will therefore be included for the sake of completeness but the comparison with other strategies that do not use the corresponding trick is unfair. To implement the genuine Dirty as described in \cite{Jalali},  we simply removed the ``$/\sqrt{\iota}$'' terms in the RevDirty script. 

\subsubsection{Practical selection of the tuning parameters}\label{sec:TuningParamSel}
All methods presented above involve tuning parameters that need to be carefully selected in practice. Generally speaking, a predefined grid of $\lambda$ values has first to be constructed. We refer to Section \ref{sec:LambdaGrid} of the Appendix for a complete description of the grid construction. Given grids of $\lambda_1$ and $\lambda_2$ values, cross-validation can be seen as the strategy of choice when  $p$ is large ({\em i.e.}, at least comparable to $n$) and/or only prediction accuracy matters. When $p\ll n$ and support recovery is of primary interest, a commonly preferred criterion is the BIC computed with unbiased estimates: for the lasso, these unbiased estimates can be obtained by the OLS-Hybrid two-step strategy \citep{LARS}, which corresponds to a simplified version of the relaxed lasso with the particular choice $\phi = 0$ \citep{M07}. For the approaches considered in this paper, it can easily be extended. This criterion will be referred to as 2stepBIC in the sequel.

\subsection{First simulation study}

\begin{figure}[h]
\begin{center}
\includegraphics[width=0.9\textwidth]{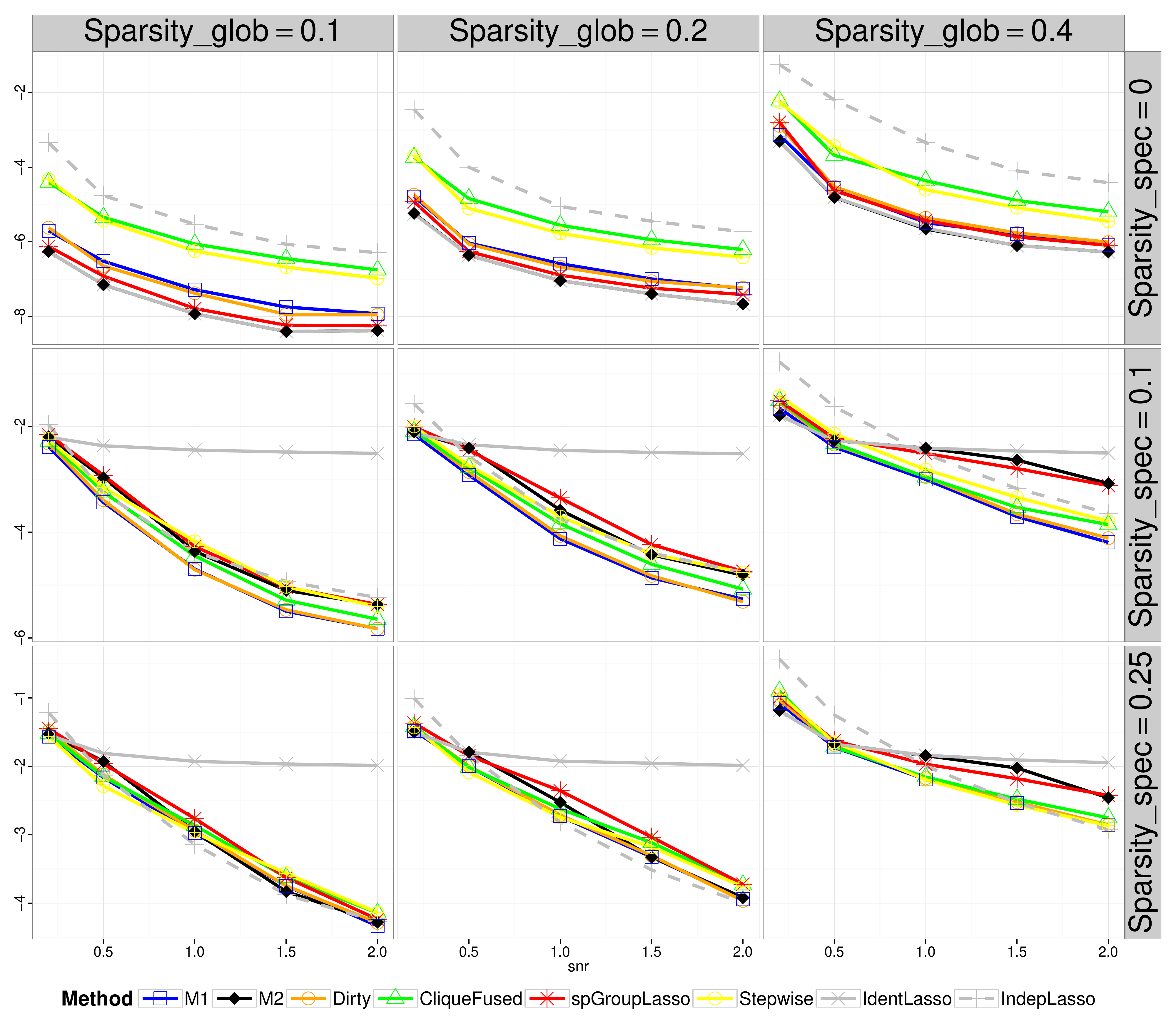}
\end{center}\caption{Results from the first simulation study: prediction accuracy.}\label{fig:L2Pred.simToep}
\end{figure}

\begin{figure}[h]
\begin{center}
\includegraphics[width=0.9\textwidth]{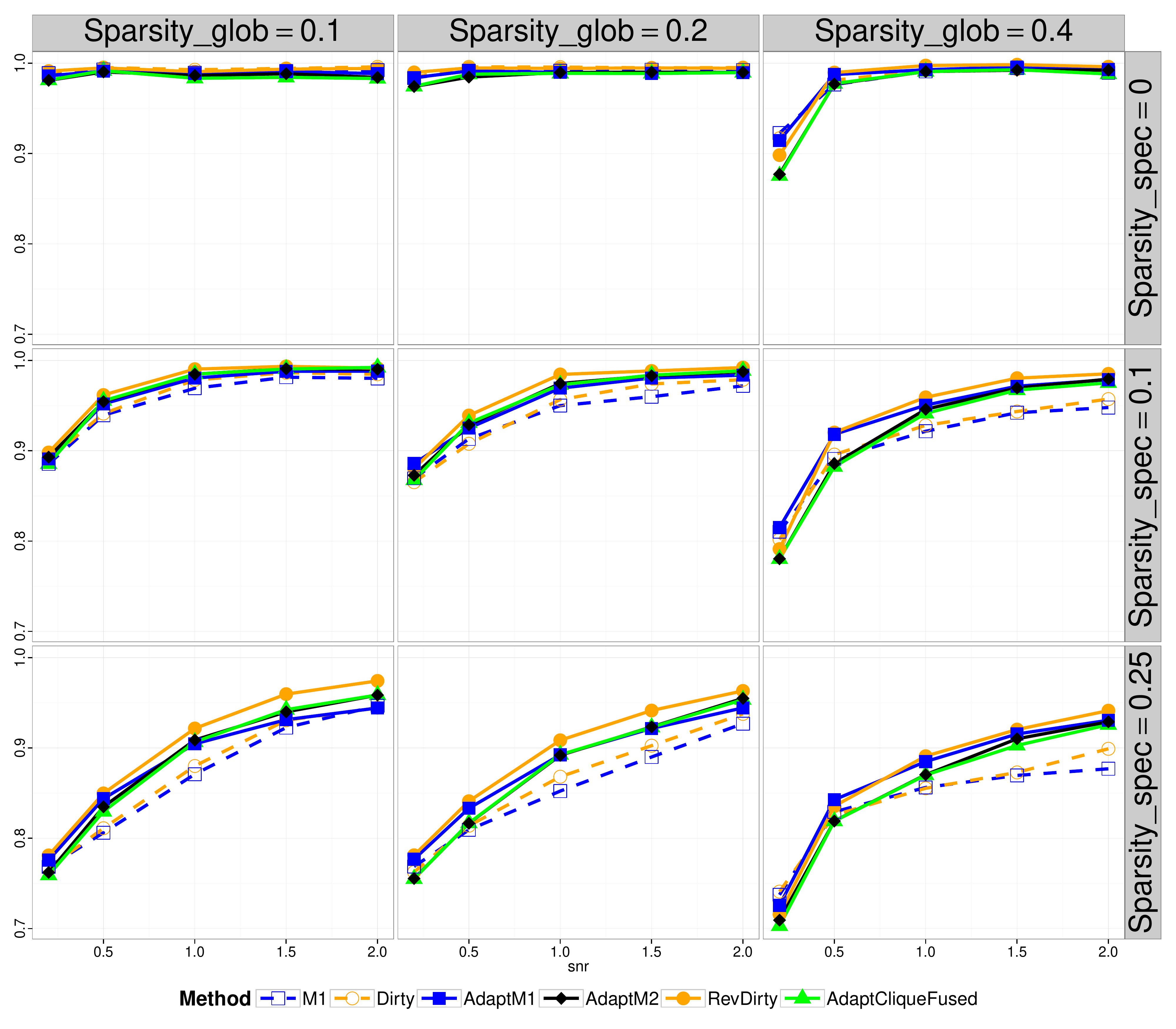}
\end{center}\caption{Results from the first simulation study: support recovery accuracy.}\label{fig:Acc.Adapt.simToep}
\end{figure}

In the first simulation study, we consider the case where $K=5$ and $n_k = 15p = 225$. For each stratum $k\in[K]$, each of the $n_k$ rows (observations) of the design matrix $\bX^{(k)}$ is generated under a multivariate Gaussian distribution ${\cal N}_p({\bf 0}_p, \boldsymbol{\Sigma})$, where the $(j_1,j_2)$-element of $\boldsymbol{\Sigma}$ is $2^{-|j_2 - j_1|}$ (for $j_1,j_2 \in[p]$). For given values $({\rm SpGlob},{\rm SpSpec})\in[0,1]^2$, each component $\betabar^*_j$ of vector $\bbetabar^*\in\R^p$ is generated from a Bernoulli distribution with parameter ${\rm SpGlob}$, while for each $k\in[K]$, each component of vector $\bdelta^*_k\in\R^p$  is generated from a Bernoulli distribution with parameter ${\rm SpSpec}$. The expected number of non-zero components in $\bbetabar^*$ and $\bdelta^*_k$ is then $p\times{\rm SpGlob}$ and $p\times{\rm SpSpec}$ respectively. Each vector $\bbeta^*_k$ are then set to $\bbetabar^* + \bdelta^*_k$, with components either 0, 1 or 2. By making SpSpec vary, we make the level of heterogeneity vary over the strata, while SpGlob enables the control of the ``common'' effects sparsity level. Observe that components $\betabar^*_j$ do not necessarily correspond to median$(\beta^*_{1,j},\ldots, \beta^*_{K,j})$ (they do in some cases, {\em e.g.}, when SpSpec$=0$, but not always). Given $\bbeta^*_k$ and $\bX^{(k)}$, vector $\bY^{(k)}$ is generated from a multivariate gaussian distribution ${\cal N}_{n_k}(\bX^{(k)}\bbeta^*_k, \sigma^2 \bI_{n_k})$. By varying the noise variance $\sigma^2$, we can make the {\em signal-to-noise-ratio} (SNR) vary. For each simulation design, corresponding to a given triple of values  $($SpGlob, SpSpec, SNR$)$, 100 replications are performed, and results presented and discussed below correspond to averages over these 100 replications. Methods are evaluated according to their prediction accuracy (Figure \ref{fig:L2Pred.simToep}), measured by $\log(\sum_{k\in[K]} \|\bX^{(k)}(\bbeta^*_k - \hbbeta_k)\|_2^2)$, and support recovery accuracy (Figure \ref{fig:Acc.Adapt.simToep}), that measures the ability to correctly recover zero and non-zero elements of matrix $\bB^* = (\bbeta^*_1,\ldots,\bbeta^*_K)$. Tuning parameters are selected with the 2StepBIC described above.

In Figure \ref{fig:L2Pred.simToep}, the first row corresponds to SpSpec$=0$, so that all $\bbeta_k^*$'s are equal. In this case, the optimal strategy is of course IdentLasso but $M_2$, spGroupLasso, $M_1$ and Dirty all perform as well (or nearly as well) as this optimal strategy. The stepwise approach and CliqueFused are less efficient, but still perform better than IndepLasso. On the other hand, when SpGlob is low and SpSpec is high (last row, first two columns), the $K$ true models have not much in common and IndepLasso is nearly-optimal: in this case again, the multi-task learning strategies perform nearly as well as the optimal strategy. In addition, and as expected, prediction performance of all methods tend to decrease as SpGlob and/or SpSpec increases, that is as the true model complexity increases. Overall, $M_1$ and Dirty show very similar performance, irrespective to the true model complexity. They perform the best for SpSpec$\neq 0$, closely followed by the other multi-task learning strategies. For SpGlob$= 0.4$ and SpSpec$\neq 0$, we observed moderate performance for both $M_2$ and spGroupLasso. As shown in Section \ref{sec:LambdaGrid} of the Appendix for method $M_2$, alternative choices for the initial grid of $(\lambda_1,\lambda_2)$ values lead to better results, suggesting that our proposal for the construction of this grid might not always be optimal for $M_2$ (and spGroupLasso) and that more theoretical effort might be needed to get a better initial grid. 

Turning our attention to support recovery, Figure \ref{fig:Acc.Adapt.simToep} presents results obtained with Dirty  and RevDirty, $M_1$, and the adaptive versions of $M_1$, $M_2$ and CliqueFused. See Figure \ref{fig:Acc.simToep} in the Appendix for the corresponding analysis of the methods compared on Figure \ref{fig:L2Pred.simToep}: overall, conclusions for these methods are very similar to those drawn from the analysis of the prediction performance. Weights for adaptive versions were derived from initial OLS estimates of the $\bbeta_k^*$. The comparison of $M_1$ with its adaptive version $M_1^{(ad)}$ clearly illustrates the potential gain on support recovery accuracy when using adaptive weights (no such gain was observed on prediction accuracy; results not shown). Moreover, adaptive versions of CliqueFused and $M_2$ attain nearly the same performance as $M_1^{(ad)}$. Overall, RevDirty performs the best, but $(i)$ the comparison is unfair since RevDirty is more than an adaptive version of Dirty and $(ii)$ adaptive versions of $M_1$ and $M_2$ performs only slightly worse in most cases.

\subsection{Second simulation study}
The objective of the second simulation study was to describe the methods performance when the dimension $p$ is of the order of $n_k$ and selection of tuning parameters is done by cross-validation. In view of the results of the first simulation study, and to save computational times, IdentLasso, IndepLasso and CliqueFused were not included in this analysis. In addition, a relaxed version of $M_1$ was included, following the idea of the relaxed lasso \citep{M07} (we made the $\phi$-value vary over the sequence $\{0, 0.1, \ldots, 0.9, 1\}$). We considered the case where $K=5$ and $p=n_k=100$. Vectors $\bbetabar$, and $\bdelta^*_k$, for $k\in [K]$ are constructed as follows. Only the first 10 components of these vectors can be non-zero. The sparsity of the common effect $\bbetabar$ is fixed to $4$, with non-zero components all equal to 1. Three levels of heterogeneity are considered: complete homogeneity (all the $\bdelta^*_k$'s are equal to ${\bf 0}_p$), low heterogeneity (the number of non-zero elements of matrix $\bDelta^*=(\bdelta^*_1,\ldots,\bdelta^*_K)$ is 5, each non-zero element being $\pm1$ with probability 1/2), and moderate heterogeneity (the overall sparsity of matrix $\bDelta^*$ is 15, each non-zero element being $\pm1$ with probability 1/2). Design matrices $\bX^{(k)}$ and response vectors $\bY^{(k)}$ are generated as in the first simulation study, except we now set $\boldsymbol{\Sigma} = \bI_p$. As in the first simulation study,  we make the SNR vary by varying the noise level.

\begin{figure}[!p]
\begin{center}
\includegraphics[width=0.9\textwidth]{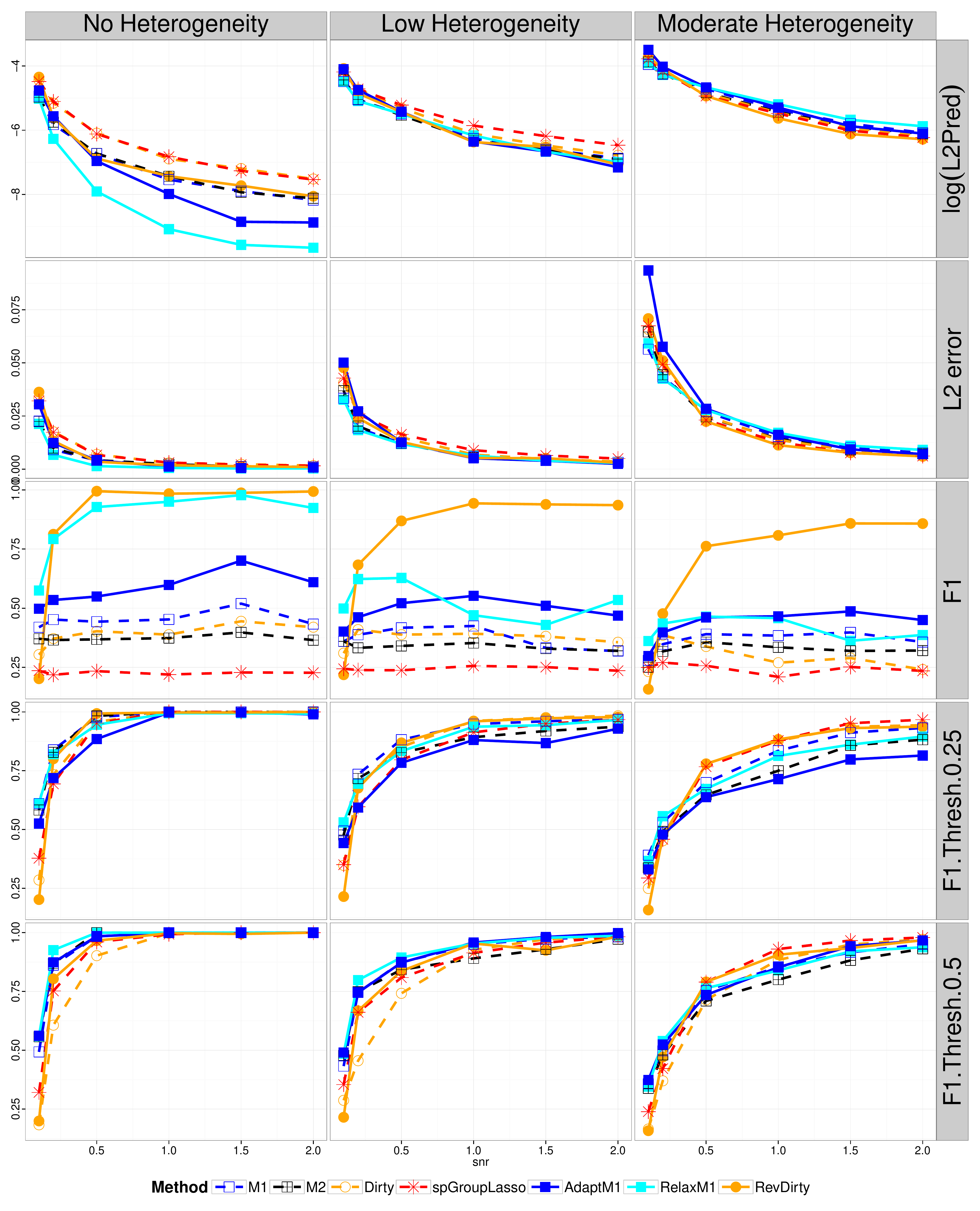}
\end{center}\caption{Results from the second simulation study with $n_k = p = 100$ and $K =5$.}\label{fig:Bigp}
\end{figure}

Results are presented in Figure \ref{fig:Bigp}. Because Dirty and RevDirty take a very long time to run in this second simulation study, results correspond to averages over 20 replications only. Methods are evaluated according to their prediction accuracy, measured by $\log(\sum_{k\in[K]} ||\bX^{(k)}(\bbeta^*_k - \hbbeta_k) ||_2^2)$, their estimation accuracy (measured by $\sum_{k\in[K]}\|\bbeta^*_k - \hbbeta_k) \|_2^2)/(Kp)$), and their ability to recover the true support (measured in this highly sparse setting by the F1-score, defined as the harmonic mean of precision and recall). Because estimators selected by cross-validation may have many tiny non-zero components, hard-thresholding these estimators can improve their performance regarding support recovery. F1-score for these hard-thresholded versions are therefore also presented, with threshold set to 0.25 and 0.5. 
By first comparing crude versions (no adaptive weights, nor relaxation nor the Revised Dirty trick), $M_1$ and $M_2$ generally perform at least similarly to Dirty and spGroupLasso. The relaxed and adaptive versions of $M_1$ generally achieve better performance than the crude $M_1$, especially in cases of complete homogeneity or low heterogeneity. In these cases, these versions also outperform RevDirty in terms of prediction and estimation accuracy. As for the F1-score, RevDirty is clearly the best: only in the case of complete homogeneity the relaxed version of $M_1$ achieves similar performance. Note however that the comparison is again unfair, since RevDirty is more than an adaptive or a relaxed version of Dirty. After the hard-thresholding step, differences among methods in terms of F1-scores are much narrower: focusing on low values of the SNR, the different versions of our approaches appear very competitive. (Our results confirm that RevDirty returns nearly unshrunk estimates and then fewer tiny non-zero components than ``purely'' $L_1$-based approaches: it does fewer mistakes (its F1-score is higher), but apparently bigger mistakes since its prediction and estimation accuracies can be outperformed by the relaxed or adaptive versions of $M_1$.)

\section{Application on road safety data}\label{sec:RealData}

One major objective in road safety is to determine factors associated with being responsible of a road traffic accident. To do so, epidemiologists typically  use data sets of car crashes, in which drivers' responsibility has been determined by experts. Of course, it is crucial that these experts use a proper and consistent ``rule'' to determine responsibility. Consistency especially implies that two experts should agree on drivers' responsibility. Level of agreement can be measured by Cohen's $\kappa$ or other standard agreement criteria, but this would require each driver's responsibility to be rated by at least two experts. Here, we will use a joint estimation strategy to assess experts agreement in situations where each driver's responsibility is only rated once by one expert. As will be made clearer below, our strategy will also help us to check the relevance of the rule used by experts for the determination of drivers' responsibility. 

Our dataset consists of $n = 5772$ road traffic accidents involving two vehicles (cars, buses, trucks,  two-wheeled motor vehicles, bikes, ...). These data have been collected in the VOIESUR project (see www.agence-nationale-recherche.fr/projet-anr/?tx\_lwmsuivibilan\_pi2[CODE]=ANR-11-VPTT-0007). For each accident, the responsibility of the two drivers was rated by one of the eighteen recruited experts. To do so, experts had to analyze police reports established after the accident. Prior to this, experts underwent a period of training where basic recommandations were given: for instance, driving under the influence (of alcohol or any other drug) should not directly be taken into account by experts, but only ``incorrect operations'' (lane departure, ...) should. This is of course critical because, otherwise, evaluating the association between alcohol consumption and being responsible of a car crash with this dataset would lead to overestimation. Experts were asked to rate responsibility from 0 (not responsible at all) to 5 (totally responsible). For the present analysis, drivers with rate 0 or 1 were considered as non-responsible, drivers with rate 4 or 5 responsible, and drivers with rate 3 (unclear responsibility) were excluded. Besides drivers' responsibility, 205 binary covariates are at our disposal, describing drivers' characteristics (age, gender, profession, alcohol and other drugs consumptions, estimated speed, lane departure, ...), their antagonists' characteristics, meteorological conditions, etc... The rule used by each expert to determine drivers' responsibility can be described by the probability of being considered as responsible by this particular expert given the 205 covariates. Then, sparse logistic regression can be used for the estimation of this probability. If experts all use the same rule, then theoretical parameter vectors $\bbeta_k^*$ of each logistic model ($k=1,\ldots,18$) should be equal. To check whether this is the case, we propose to jointly estimate the 18 corresponding sparse logistic models using the two approaches described in this paper (both the crude and refined versions). Each method returns a $(206,18)$ matrix where each columns contains the parameter vector (including the intercept) of one stratum, {\em i.e.} one expert. We evaluated both methods on a predefined grid of $(\lambda_1,\lambda_{2,k})$ values and retained the model minimizing the 2stepBIC. Contrary to situations considered in our simulation study, the $n_k$'s are not constant here and several strategies can be put forward for determining the grid. The most simple one consists in using $\lambda_{2,k} = \lambda_2$ and proceed as described in Section \ref{sec:TuningParamSel}. As mentioned in Section \ref{sec:TuningParamSel} (and as suggested by the preliminary non-asymptotic analysis of Section \ref{sec:NonAsympResults} of the Appendix),  another strategy for method $M_1$ consists in first standardizing columns of matrix $\bcX$, then setting $\lambda_{2,k} = \lambda_2$ and finally proceeding again as described in Section \ref{sec:TuningParamSel}. Both strategies were applied for method $M_1$ and they led to similar results.

With either method, good concordance was observed between supports of vectors $\hbbeta_k$: only two [resp. three] elements of matrix $\widehat\bDelta = (\hbdelta_1,\ldots,\hbdelta_{18})$ were non-null when using $M_1$ [resp. $M_2$]. Moreover, non-zero elements returned by $M_2$ have small absolute values ($\leq $ 0.2). This suggests good agreement among the 18 experts and we can consider they use very similar rules (if not equal) to determine drivers' responsibility. An interesting result is that elements in $\widehat\bDelta$ related to driving under the influence were all zero: this suggests that all experts account for alcohol consumption in the same way. 

Now, looking at the common vector $\hbbetabar$ can further help us to check the adequacy of the common rule used by the experts. We will pay a particular attention on the association between alcohol consumption and the probability of being considered responsible by experts. Indeed for road safety matters, and as recalled to experts during the training phase, the only fact of driving under the influence should not imply to be (nor increase the risk of being) considered as responsible. If components of vectors $\hbbetabar$ corresponding to alcohol consumption are non-null, then the adequacy of the rule may be questioned.  Method $M_1$ returns  a vector $\hbbetabar$
with 74 non-null components, while $M_2$ returned a slightly sparser one with 60 non-null components. Overall, variables that are the most associated to a high probability of being responsible are as expected: a driver who drove (or ran) away, and/or went through a stop sign or a red light, crossed the lane, etc. is likely to be considered as responsible of the accident. Regarding alcohol consumption, both $M_1$ and $M_2$ return two non-zero components in vector $\hbbetabar$: those corresponding to the highest level of alcohol (between $1.2$ and 2 g/L, and above 2g/L respectively). Two interpretations can be put forward. A pessimistic one is that experts did not follow recommandations and usually considered these levels of alcohol as indicators of drivers' responsibility. Another more optimistic interpretation is that drivers presenting these levels of alcohol may have behaviors that can not be captured  by a simple linear combination of the 205 available binary variables (either because interactions should be considered or, more likely, because some descriptors are missing). Choosing between these two interpretations is out of the scope of the present analysis and will be the focus of future work. 

Results from our approaches were compared to a more standard strategy in epidemiology which consists in using the SAS LOGISTIC procedure along with STEPWISE option to select significant covariates (including interaction terms between the 205 original covariates and indicator functions representing each expert). The STEPWISE procedure was not able to run with the whole set of interaction terms. We then had to focus on the 205 original covariates plus interaction terms involving alcohol-related covariates. The results returned by SAS were consistant with those obtained with our approaches. More precisely, no interaction term was retained by the STEPWISE procedure confirming that all experts account for alcohol in the same way when determining drivers' responsibility. Moreover, the STEPWISE retained terms corresponding to the two highest levels of alcohol, as our approaches did. This example illustrates situations where our approaches, though very simple to implement, extend the scope of standard procedures used in epidemiology.

\section{Discussion}\label{sec:Discussion}

In this paper we present a new approach, along with refinements, aimed at jointly modeling several related regression models. Although most of the presentation was made in the situation where these models correspond to several strata of a population, our approach can be used in the more general multi-task learning setting. Also, for ease of notation mostly linear regression models were considered in this paper. As illustrated in the application of Section  \ref{sec:RealData}, extension to other models is however straightforward. This is particularly true for our first approach and its adaptive or relaxed version, since they all can be rewritten as simple weighted lasso problems: in particular, linear, logistic, Poisson and Cox models can be treated thanks to the glmnet R package \citep{glmnet}. Other functions exist for other models, like the clogitL1 R package for conditional logistic models of \cite{cLogitL1}, and allow for further extensions of our approach.

Our approach naturally connects with several strategies formerly proposed in the literature. First, it is a simplified version of the Dirty models proposed by \cite{Jalali} in the linear regression setting, and is much easier to extend to other regression models. On the designs considered in our simulation study, it still performs similarly to the Dirty models and we also exhibited situations where its theoretical sample complexity is at least as good as that of the Dirty models. Second, adaptive versions of our approaches are special cases of the generalized fused lasso. This connection is particularly appealing for interpretation matters since it allows for a simple comparison between CliqueFused, the InterLasso and our two strategies. First, CliqueFused uses cliques, while InterLasso and our two approaches use star-graphs. Second, for all $j\in[p]$, the center of the star-graphs are set to $\beta_{\ell,j}$ when using InterLasso with reference stratum $\ell$, while when using adaptive versions of our approaches centers are adaptively set to $\beta_{\ell_j,j}$: the reference stratum is covariate-specific, automatically selected from the initial estimates, and it corresponds (with probability tending to 1) to one of the stratum in $K^*_j = \{k: \beta^*_{k,j} = \beta^*_{\ell^*_j,j}\}$ if the initial estimates are consistent. Finally, besides the structure of the graph (edges corresponding to differences $\beta_{k,j}-\beta_{k',j}$ that are penalized), these four strategies exhibit differences according to which nodes in the graph (coefficients $\beta_{k,j}$) are directly encouraged to be sparse: they are all penalized when using CliqueFused or $M_2$, while only the $\beta_{\ell,j}$ for InterLasso and the $\beta_{\ell_j,j}$ for $M_1$ are penalized, for $j\in[p]$. 


From a theoretical point of view, the connection with the generalized fused lasso allowed us to derive asymptotic oracle properties for our adaptive versions. Preliminary non-asymptotic results were also stated, under strong but not necessary conditions. Regarding prediction performance, these preliminary results might be extended to more general settings by adapting the recent results of \cite{Dalalyan14}, who study prediction performance of the lasso in the presence of correlated designs. As for support recovery of the $\bbeta^*_k$'s, adapting the dual-witness idea of \cite{Wainwright2009} is a promising lead (as \cite{Jalali} did for the Dirty models). Other extensions of this work could rest on greedy algorithms or recent MCMC-based approaches (such as the exponential screening of \cite{RigolletTsyb}), for which optimal prediction bounds have been obtained with no particular assumption on the design matrix.


\newpage

\section{Appendix: Supplementary Materials}

\subsection{Theoretical results}\label{Theo_Supp}

In this section, we derive preliminary non-asymptotic properties for our first approach $M_1$ (Sections \ref{sec:SampleComp} and \ref{sec:NonAsympResults}) as well as asymptotic oracle properties for our adaptive versions (especially for $M_2$, see Section \ref{Asymtpot}).

\subsubsection{Sample complexity under independent Gaussian designs, when $K=2$ and $\beta_{1,j}^*\beta_{2,j}^*\geq 0$ for all $j\in[p]$.}\label{sec:SampleComp}
Here we consider the case where $K=2$. Working under the assumptions considered in \cite{Jalali}, it is easy to show that the sample complexity of $M_1$ is at most the one of the Dirty models, under the additional assumption that $\beta_{1,j}^*\beta_{2,j}^*\geq 0$ for all $j\in[p]$. Indeed, for any given pair of potential estimates $\beta_{1,j}$ and $\beta_{2,j}$, the penalty term involved in the Dirty approach  for covariate $j$ can be written 
$$ \phi(r_1,r_2) = \lambda_1 \max(|r_1|,|r_2|) + \lambda_2 \{|\beta_{1,j} - r_1| + |\beta_{2,j} - r_2|\},$$
while for $M_1$ it reduces to  
$$  \bar \phi(r) = \lambda_1 |r| + \lambda_2 \{|\beta_{1,j} - r| + |\beta_{2,j} - r|\} = \phi(r,r).$$
Moreover, for any positive $\lambda_1,\lambda_2$ and any $\beta_{1,j},\beta_{2,j}$ it is easy to show that
$\min_{r_1,r_2}\phi(r_1,r_2) = \min_r \bar \phi(r)$ if $\beta_{1,j}\beta_{2,j}\geq 0$ 
and $\min_{r_1,r_2}\phi(r_1,r_2) < \min_r \bar \phi(r)$ otherwise. In words, pairs of values such that $\beta_{1,j}\beta_{2,j}\geq 0$ are more heavily penalized by $M_1$, while all other pairs are equally penalized by Dirty and $M_1$. Consequently, assuming that  $\beta^*_{1,j}\beta^*_{2,j}\geq 0$  for all $j\in[p]$, the  probability of correct support recovery for $M_1$ is superior or equal to that of Dirty. In particular, under the assumptions considered in Theorem 3 of \cite{Jalali}, and assuming that $\beta^*_{1,j}\beta^*_{2,j}\geq 0$ for all $j\in[p]$, perfect support recovery is ensured for method $M_1$ as soon as the common number of observations per stratum (task) is superior to $(2-\alpha)s\log(p - (2-\alpha)s)$, where $s$ is the common support size of $\bbeta^*_{1}$ and $\bbeta^*_{2}$ and $\alpha$ is the overlap proportion between the two supports. As shown in \cite{Jalali} for their dirty model, this implies that $M_1$ outperforms both IndepLasso and the $L_1/L_\infty$-group lasso strategy in terms of sample complexity (except in the two extremes situations of no sharing at all and full sharing, where it matches the best strategy).

\subsubsection{Other preliminary non-asymptotic properties}\label{sec:NonAsympResults}

In addition to the sample complexity in the simple case described above, other preliminary non-asymptotic properties can easily be derived for our approach thanks to its rewriting as a weighted lasso. Most non-asymptotic theoretical results of the lasso have been established under strong conditions on the design matrix: to name a few, the irrepresentability condition, the mutual incoherence, the restricted eigenvalue condition or the RIP property (see \cite{vdGBEJS} for an overview). In Equation (\ref{eq:M1_Lasso}) of the main paper, it is clear that matrix $\bcX$ generally does not enjoy such properties. Assuming for simplicity that $n_k = n/K$ for all $k\in[K]$, we present here simple situations where it does. A more thorough study would be needed to establish non-asymptotic properties under more general assumptions and an interesting lead lies in the recent results concerning prediction accuracy of lasso estimators under correlated designs \citep{Dalalyan14}. This will be considered elsewhere.

The first situation is when $K$ is large enough and the $\bbeta^*_k$'s remain highly similar. More precisely, assume that  design matrices ${\bX^{(k)}}$ fulfill the following mutual incoherence condition \citep{LouniciLasso}, for all $k\in[K]$: setting  
$\Sigma^{(k)} = ({\bX^{(k)}}^T\bX^{(k)})/n_k $ for all $k\in[K]$, we assume that $\Sigma^{(k)}_{j,j} =1$ for all $j\in[p]$ and $\max_{j_1\neq j_2} |\Sigma^{(k)}_{j_1,j_2}| \leq 1/(7\alpha s)$, for some integer $s\geq 1$ and constant $\alpha>1$. Then the following standardized version of matrix $\bcX$
$$  
\widetilde\bcX = \left( 
\begin{array}{c c c c }
\bX^{(1)} &\sqrt{K}\bX^{(1)}& \hdots &{\bf 0}  \\
\vdots & \vdots & \ddots & \vdots \\
\bX^{(K)} &{\bf 0}&\hdots  & \sqrt{K} \bX^{(K)}  
\end{array}
\right)
$$ 
can easily be shown to fulfill the same mutual incoherence assumption, as long as $K > (7\alpha s)^2$. For such values of $K$, and assuming that the true parameter vector $\btheta^*$ contains at most $s$ non-zero components (such a $\btheta^*$ is ensured to be unique), results of \cite{LouniciLasso} can be applied to derive the $L_\infty$ rate of convergence and the sign consistency of estimators obtained from $M_1$, with sparsity parameters set to $\lambda_{2,k}=\lambda_1$ (which is equivalent to working with the non-standardized matrix $\bcX$ and $\lambda_{2,k} = \lambda_1 K^{-1/2}$). Of course the assumption $K > (7\alpha s)^2$ is very strong: it holds typically when strata are numerous and highly similar (so that $s$ remains low). 

Another simple situation arises when ${\bX^{(k)}}^T\bX^{(k)}/n_k = \bI_p$ for all $k\in[K]$ and vectors $\bbeta^*_k$ are all equal, and we set again $\lambda_{2,k} = \lambda_1$. Setting $J_0=\{j: \theta_j \neq 0\}$, this ensures that $J_0 \subseteq [p]$, i.e. $\btheta^*_{J_0}  = \bbetabar^*_{J_0}$. Then it is easy to show that the irrepresentability condition of \cite{Wainwright2009} holds with $\gamma = 1-\sqrt{1/K}$, and then to obtain the following $L_\infty$ bound: for any $\delta,\varepsilon>0$, and assuming that the noise variables $\xi^{(k)}_i$ are i.i.d. $\sigma$-sub-Gaussian, we have 
$$ \max_k \| \hbbeta_k - \bbeta_k^*\|_\infty = \|\hbbetabar - \bbetabar^*\|_\infty \leq \sigma \Bigg\{ \sqrt{\frac{2 \log |J_0| + \varepsilon^2}{n}} + \frac{2}{\gamma}\sqrt{\frac{2|J_0|\log(Kp - |J_0|) + \delta^2}{n}} \Bigg\}, $$ 
with probability greater than $1 -2\exp(-\delta^2/2) - 2\exp(-2\varepsilon^2/2)$. 
This bound, combined with a ``beta-min'' condition, leads to the sign consistency of the approach. It is noteworthy that the IdentLasso would lead to the same kind of bound, with an only slightly better second term: $ 2\sqrt{\{2|J_0|\log(p - |J_0|) + \delta^2\}/n}$. In other words, when ${\bX^{(k)}}^T\bX^{(k)}/n_k = \bI_p$ for all $k\in[K]$ and all the $\bbeta^*_k$'s are equal our approach performs as well as IdentLasso (up to constants and $\log$-terms), which is optimal in this case. 

\subsubsection{Asymptotic results for the adaptive version of $M_2$ in the fixed $p$, fixed $K$ case}\label{Asymtpot}

In this paragraph, we will consider the simple situation where $p$ and $K$ are held fixed (they do not increase with $n$). We further assume that, for all $k\in[K]$, matrices ${\bX^{(k)}}^T \bX^{(k)}/n_k$ converges to positive definite matrices $\bC^{(k)}$ as $n_k\rightarrow \infty$, and  $n_k/n\rightarrow \kappa_k$ as $n\rightarrow \infty$, with $0<\kappa_k<1$ (i.e., {\em stratum} sizes all tend to infinity at the same rate). These assumptions essentially imply that MLEs $\widetilde \beta_{k,j}$ can be used in the definition of the weights in Equations (\ref{est_adaM2}) and (\ref{est_adaM1}) of the main paper, and are $\sqrt{n}$-consistent as $n\rightarrow\infty$.

Before stating our result, some notations are needed; in particular, the overall complexity of the model returned by $M_2^{ad}$ has to be defined properly. For any $j\in[p]$, introduce the quantities $\ell_j^* = \min\{k: \beta^*_{k,j} \in{\rm median}(\beta^*_{1,j},\ldots,\beta^*_{K,j})\}$, $K^*_j = \{k: \beta^*_{k,j}=\beta^*_{\ell_j^*,j}\}\subseteq [K]$, and $N^*_j = \sum_{k\in \ell^*_j}n_k$. Further introduce ${\cal A}^* = \{(k,j): \beta^*_{k,j}\neq 0\}$ and, for all $j\in [p]$, ${\cal A}_j^* = \{k: \beta^*_{k,j}\neq 0\}$ and $K^*_{{\cal A}^*,j} = \{k: \beta^*_{k,j}=\beta^*_{\ell_j^*,j}\neq 0\}$. 

To define the empirical counterparts of the previous quantities returned by $M_2^{ad}$, first recall that  for any $j\in[p]$,   $\ell_j = \min\{k: \widetilde\beta_{k,j} \in{\rm median}(\widetilde\beta_{1,j},\ldots,\widetilde\beta_{K,j})\}$. Then denote by $\hbeta^{(ad)}_{k,j}$ for any $k\in [K]$ and $j\in[p]$ the estimator of $\beta^*_{k,j}$ returned by $M_2^{ad}$. We can now define $\widehat {\cal A} = \{(k,j): \hbeta^{(ad)}_{k,j}\neq 0\}$, and, for all $j\in [p]$, $\widehat \ell_j = \{k: \hbeta^{(ad)}_{k,j}=\hbeta^{(ad)}_{\ell_j,j}\}\subseteq [K]$,  $\widehat {\cal A}_j = \{k: \hbeta^{(ad)}_{k,j}\neq 0\}$  and $\widehat K_{\widehat{\cal A},j} = \{k: \hbeta^{(ad)}_{k,j}=\hbeta^{(ad)}_{\ell_j,j}\neq 0\}$. 
Finally for any $j\in[p]$ denote by $\widehat s_j$ the number of  distinct non-null values in $(\hbeta^{(ad)}_{1,j},\ldots,\hbeta^{(ad)}_{K,j})$, that is
$$\widehat s_j =\left\{
\begin{array}{ll} 
|\widehat {\cal A}_j| & {\rm if }\ \hbeta^{(ad)}_{\ell_j,j}=0\\
|\widehat {\cal A}_j| - |\widehat K_{\widehat{\cal A},j}| + 1& {\rm otherwise}.
\end{array}\right.$$ 
The overall complexity of the model returned by $M_2^{ad}$, which corresponds to the number of ``free'' parameters returned by $M_2^{ad}$, is then defined as $\widehat s = \sum_{j\in[p]}\widehat s_j$. More precisely, for any $j\in[p]$ such that $\widehat s_j>0$, $M_2^{ad}$ returns a vector $\widehat{\boldsymbol \eta}^{(ad)}_j$ of size $\widehat s_j$ defined by    
$$
\widehat{\boldsymbol \eta}^{(ad)}_j =\left\{
\begin{array}{ll}
\hbeta^{(ad)}_{\ell_j,j}       &{\rm if }\ \hbeta^{(ad)}_{\ell_j,j} \neq 0\ {\rm and }\ \widehat  s_j=1;\\
(\hbeta^{(ad)}_{\bar k_{j,1},j}, \ldots, \hbeta^{(ad)}_{\bar k_{j,\widehat s_j},j})^T       &{\rm if }\ \hbeta^{(ad)}_{\ell_j,j} = 0\ {\rm and }\ \widehat s_j\geq 1, \ {\rm with}\ \{\bar k_{j,1}, \ldots, \bar k_{j,\widehat s_j}\} =\widehat  {\cal A}_j;\\
(\hbeta^{(ad)}_{\ell_j,j}, \hbeta_{\bar k^{(ad)}_{j,1},j}, \ldots, \hbeta^{(ad)}_{\bar k_{j,\widehat s_j -1},j})^T      &{\rm otherwise, \ with}\ \{\bar k_{j,1}, \ldots, \bar k_{j,\widehat s_j -1}\} = \widehat {\cal A}_j \setminus \widehat K_{{\cal A},j}.\\
\end{array}\right.
$$

Now denote by $\widehat {\boldsymbol \eta}^{(ad)}$ the estimator returned by $M_2^{ad}$ (after some reordering), {\em i.e.},  $\widehat {\boldsymbol \eta}^{(ad)}= (\widehat{\boldsymbol \eta}^{(ad)}_j )_{j: \widehat s_j>0}\in\R^{s}$. We can further introduce the theoretical counterpart of $\widehat s_j$ and $\widehat s$, by setting $s^*_j = |{\cal A}_j^*|$ if 
$\beta^*_{\ell^*_j,j}=0$ and $s^*_j =|{\cal A}^*_j| - |K^*_{{\cal A}^*,j}| + 1$ otherwise, and $s^*=\sum_{j\in[p]} s^*_j$. Finally introduce 
for any $j\in[p]$ such that $s^*_j>0$,   
$$
{\boldsymbol \eta}^*_j =\left\{
\begin{array}{ll}
\beta^*_{\ell^*_j,j}       &{\rm if }\ \beta^*_{\ell^*_j,j} \neq 0\ {\rm and }\ s^*_j=1;\\
(\beta^*_{\bar k^*_{j,1},j}, \ldots, \beta_{\bar k^*_{j,s_j},j})^T       &{\rm if }\ \beta^*_{\ell^*_j,j} = 0\ {\rm and }\ s^*_j\geq 1, \ {\rm with}\ \{\bar k^*_{j,1}, \ldots, \bar k^*_{j,s_j}\} = {\cal A}^*_j;\\
(\beta^*_{\ell^*_j,j}, \beta^*_{\bar k^*_{j,1},j}, \ldots, \beta^*_{\bar k^*_{j,s^*_j -1},j})^T      &{\rm otherwise, \ with}\ \{\bar k^*_{j,1}, \ldots, \bar k^*_{j,s^*_j -1}\} = {\cal A}^*_j \setminus K^*_{{\cal A}^*,j}.\\
\end{array}\right.
$$
and ${\boldsymbol \eta}^*= ({\boldsymbol \eta}^*_j )_{j: s^*_j>0}\in\R^{s^*}$.

\begin{figure}
\begin{center}
\includegraphics[width=0.99\textwidth]{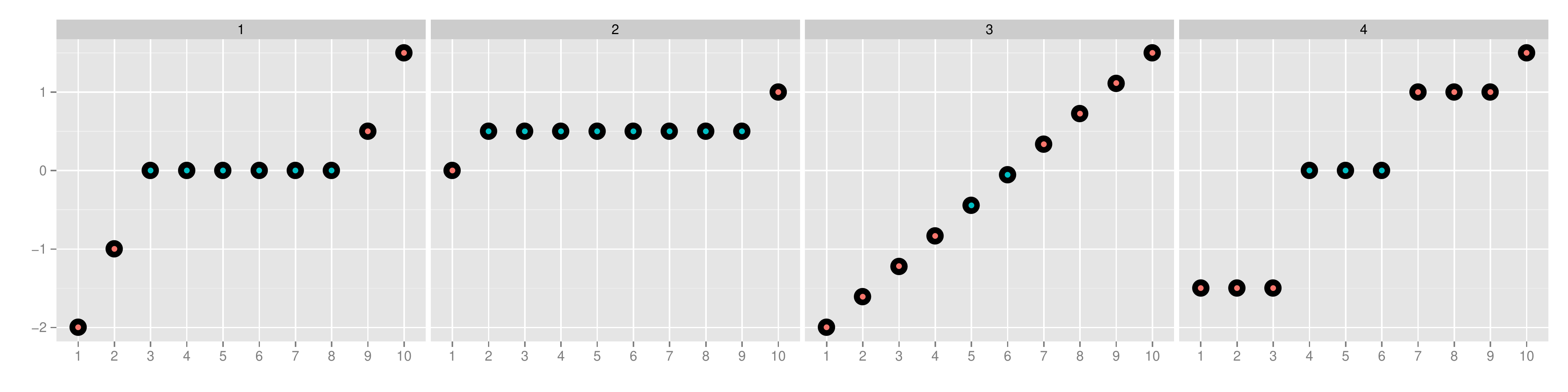}
\end{center}\caption{Graphical representation of four typical examples of vector $(\beta^*_{1,j},\ldots,\beta^*_{K,j})$, for some $j\in[p]$ and with $K=10$. Without loss of generality, we assume here that strata $\{1,\ldots, K\}$ are ordered in such a way that $\beta^*_{1,j}\leq\ldots\leq\beta^*_{K,j}$. For each example, the median value of $(\beta^*_{1,j},\ldots,\beta^*_{K,j})$ is represented by a blue bullet: under the third example, all values between the two blue bullets $[-0.45,0]$ are valid median values, while only one value corresponds to the median under the other three examples (0, 0.5 and 0 for examples 1, 2 and 4 respectively).}\label{fig:confbetas}
\end{figure}

As a simple consequence of Theorem 2 in \cite{Viallon}, we have $\P(\widehat {\cal A}={\cal A}^*)\rightarrow 1$ and $\P(\cap_{j\in[p]}\{ \widehat K_{{\cal A},j} = K^*_{{\cal A}^*,j}\} )\rightarrow 1$, as $n\rightarrow\infty.$ In particular, this implies that the two vectors ${\boldsymbol \eta}^*$ and $\widehat{\boldsymbol \eta}^{(ad)}$ are of identical size $s^*=\widehat s$ with probability tending to one. Moreover, we have the following Gaussian limit distribution 
$$ \sqrt{n}(\widehat{\boldsymbol \eta}^{(ad)} - {\boldsymbol \eta}^*) \rightarrow {\cal N}({\bf 0}_{s^*}, \sigma^2 (\widetilde{\cal X}_{s^*}^T\widetilde{\cal X}_{s^*})^{-1})\quad{as}\ n\rightarrow \infty,$$
with $\widetilde{\cal X}_{s^*}$ the $(n\times s^*)$ matrix that can be written as $(\widetilde{\cal X}_{j,s^*_j})_{j:s^*_j>0}$ where each submatrix $\widetilde{\cal X}_{j,s^*_j}$, of size $(n\times s^*_j)$ is of the form:
$$
\widetilde{\cal X}_{j,s^*_j}=\left\{
\begin{array}{ll}
\displaystyle\sum_{k\in K^*_{{\cal A}^*,j}} \widetilde X^{(k)}_j     &{\rm if }\ \beta^*_{\ell^*_j,j} \neq 0\ {\rm and }\ s^*_j=1;\\
(\widetilde X^{(\bar k^*_{j,1})}_{j}, \ldots, \widetilde X^{(\bar k^*_{j,s^*_j})}_{j})       &{\rm if }\ \beta^*_{\ell^*_j,j} = 0\ {\rm and }\ s^*_j\geq 1, \ {\rm with}\ \{\bar k^*_{j,1}, \ldots, \bar k^*_{j,s_j}\} = {\cal A}^*_j;\\
\displaystyle(\sum_{k\in K^*_{{\cal A}^*,j}} \widetilde X^{(k)}_j , \widetilde X^{(\bar k^*_{j,1})}_{j}, \ldots, \widetilde X^{(\bar k^*_{j,s_j})}_{j})      &{\rm otherwise, \ with}\ \{\bar k^*_{j,1}, \ldots, \bar k^*_{j,s^*_j -1}\} = {\cal A}^*_j \setminus K^*_{{\cal A}^*,j}.\\
\end{array}\right.
$$
Here, $\widetilde X_j^{(k)}$ is the column vector of size $n$ of the form $({\bf 0}^T_{\sum_{k'<k}n_{k'}},{X_j^{(k)}}^T,{\bf 0}^T_{\sum_{k'>k}n_{k'}})^T$, with $X_j^{(k)}$ the $j$-th column of matrix $\bX^{(k)}$. For any $j\in[p]$, if $\beta_{\ell^*_j,j}^*\neq 0$ then we have $\beta^*_{k,j}=\beta_{\ell^*_j,j}^*$ for all $k\in  
K^*_{{\cal A}^*,j}$ and these parameters are all estimated by the common value $\hbeta^{(ad)}_{\ell_j,j}$, which has the same Gaussian limit distribution as that of the estimator we would obtain by pooling all data corresponding to strata in $K^*_{{\cal A}^*,j}$ together. Our results then show that method $M_2^{ad}$ enjoys {\em asymptotic oracle properties}. More precisely, method $M_2^{ad}$ is asymptotically optimal in situations where, for all $j\in[p]$, $\beta^*_{k_1,j}=\beta^*_{k_2,j}$ implies that either $\beta^*_{k_1,j}=\beta^*_{k_2,j}=0$ or $\beta^*_{k_1,j}=\beta^*_{k_2,j}=\beta^*_{\ell_j^*,j}$, as in examples 1 to 3 of Figure \ref{fig:confbetas} of this Appendix. Asymptotic oracle properties can easily be derived for method $M_1^{ad}$ as well but, for $M_1^{ad}$ to be optimal we must have in addition: for all $j\in[p]$, $\{\beta^*_{\ell^*_j,j}\neq 0\} \Rightarrow \{\forall k\in[K], \beta^*_{k,j}\neq 0\}$. For instance, in example 2 on Figure \ref{fig:confbetas}, the estimator of $\beta^*_{1,j}$ returned by $M_1^{ad}$ can not be 0 (unless $M_1^{ad}$ returns a zero estimate for $\beta^*_{\ell_j,j}$ as well), so that $M_1^{ad}$ would typically return a model with overall complexity higher than the theoretical one.  On the other hand, contrary to the CliqueFused strategy that is based on cliques, both our approaches use star-graphs and are therefore sub-optimal in situations where non-zero values in $(\beta^*_{1,j},\ldots,\beta^*_{K,j})$ consist of at least two groups of identical values (and possibly some distinct non-zero values). For instance, in example 4 of Figure \ref{fig:confbetas}, neither $M_1^{ad}$ nor $M_2^{ad}$ can return equal values for components $\hbeta^{(ad)}_{1,j}$, $\hbeta^{(ad)}_{2,j}$, and $\hbeta^{(ad)}_{3,j}$ (nor for $\hbeta^{(ad)}_{8,j}$, $\hbeta^{(ad)}_{9,j}$, and $\hbeta^{(ad)}_{10,j}$), while CliqueFused would, for $n$ large enough. However, CliqueFused, {\em i.e.}, clique-based strategies, can of course be outperformed on finite samples. This was illustrated in the simulation study conducted in \cite{Viallon} in the single-stratum setting. In the present ``multiple strata'' setting, if $\beta^*_{1,j},\ldots,\beta^*_{K-1,j}$ are all equal and $\beta^*_{K,j}$ is different from the $K-1$ other ones, we would have $\ell_j\in[K-1]$ for $n$ enough, and our approaches would only penalize $|\beta_{K,j} - \beta_{\ell_j,j}|$: they are therefore more likely to detect this difference than the approach based on clique-graphs which penalizes all the differences $|\beta_{K,j} - \beta_{k,j}|$ for all $k\in[K-1]$.

\subsection{Algorithm for method $M_2$}\label{sec:ImplementationM2}
In this paragraph, we consider a general framework that encompasses both linear and logistic models along with SVM as special cases. Let $f$ be a convex loss function and for any $k\in[K]$ set
$\LL_k(\bbeta_{k}) =\sum_{i\in [n_k]}f(\bbeta_{k}^T\ba^{(k)}_i 
+c^{(k)}_i)$ for $\bbeta_k\in\R^p$ and
some given $\ba^{(k)}_i \in\R^{p}$ and $c^{(k)}_i\in\R$. We define the
$k$-th {\em feature matrix} $\bA^{(k)} :=[\ba^{(k)}_1,\ldots,\ba^{(k)}_{n_k}]^T\in\R^{n_k \times p}$. We consider a generic supervised learning problem of the form
\begin{eqnarray}
({\hbbetabar},\dgamman) &\in&
\argmin_{{\bbetabar},\dgamma} \:
\sum_{k=1}^K \Big\{\LL_k({\bbetabar}+\bgamma_{k}) 
+ \lambda_{1,k} \|{\bbetabar}+\bgamma_{k}\|_1 + \lambda_{2,k} \ \|\bgamma_{k}\|_1\Big\}.
\label{eq:lin_model_M5_add_general}
\end{eqnarray}

Recall that throughout the paper, data is of the
form $(\bx^{(k)}_1,y^{(k)}_1),\ldots,(\bx^{(k)}_{n_k},y^{(k)}_{n_k})$ with $\bx^{(k)}_i\in\R^p$ and
$y^{(k)}_i\in\R$, $i\in[n_k]$ and $k\in[K]$. In this context, the aforementioned
formalism covers the linear regression with the loss function set to the squared loss: $f(\xi) = f_{\rm
sq}(\xi)= \xi^2/(2n)$, and by setting $\ba^{(k)}_i=\bx^{(k)}_i\in\R^p$  the
vector of covariates and $c^{(k)}_i=-y_i^{(k)}$ the (negative) response for observation $i\in[n_k]$ of the $k$-th stratum.
Likewise, by setting
$f(\xi)=f_{\rm log}(\xi)=\log(1+e^{-\xi})$, $c_i^{(k)}=0$, and $\ba^{(k)}_i=y^{(k)}_i\bx^{(k)}_i$ for $i\in[n_k]$,
our formalism covers both the logistic regression and the SVM framework, by considering the logistic loss $f(\xi)=f_{\rm log}(\xi)=\log(1+e^{-\xi})$ and the hinge loss $f(\xi)=f_{\rm hi}(\xi) = (1-\xi)_+$ respectively. \vskip5pt

For future use, we denote by $f^\ast$ the Fenchel conjugate of the
loss function $f$, which is the extended-value convex function
defined as
\[
f^\ast(\vartheta) := \max_\xi \: \xi\vartheta - f(\xi).
\]
Beyond convexity, we make a few mild assumptions about the loss
function $f$. First, we assume that it is non-negative everywhere,
and that it is closed (its epigraph is closed), so that
$f^{\ast\ast} = f$. These assumptions are met with the squared,
logistic and hinge loss functions, as well as other popular loss
functions. The conjugate of the squared,
logistic and hinge loss functions are given in
Table~\ref{tab:conj-loss}.

\begin{table}[t]
\centering
\begin{tabular}{r|c|c|c}
& loss function $f(\xi)$ & conjugate function $f^\ast(\vartheta)$ & domain of $f^\ast$ \\[.05in]\hline
\rule{0pt}{2.6ex} squared & $ f_{\rm sq}(\xi) = \xi^2/(2n)$ & $ \vartheta^2/(2n)$ & $\R$ \\
logistic & $f_{\rm log}(\xi) = \log(1+e^{-\xi})$ &
$(-\vartheta)\log(-\vartheta) + (\vartheta+1) \log(\vartheta+1)$ & $[-1,0]^m$ \\
hinge & $f_{\rm hi}(\xi) = (1-\xi)_+$ & $-\vartheta$ & $[-1,0]^m$
\end{tabular}
\caption{Expression for the conjugate of popular loss functions,
adopting the convention $0\log 0 = 0$ for the logistic
loss.}\label{tab:conj-loss}
\end{table}

We have
\begin{eqnarray*}
\phi((\lambda_{1,k})_{k\in[K]},(\lambda_{2,k})_{k\in[K]})&=&
\min_{\bbetabar,\dgamma} \:
\sum_{k=1}^K \left\{\LL_k({\bbetabar}+\bgamma_{k}) + \lambda_{1,k} \|{\bbetabar}+\bgamma_{k}\|_1 +\lambda_{2,k} 
\|\bgamma_{k}\|_1\right\} \\
&=& \min_{\substack{\bbetabar,\dbeta,\\\zz}}
\sum_{k=1}^K \left\{G_k(\bz_k) + \lambda_{1,k} \|{\bbeta_k}\|_1
+\lambda_{2,k} \|\bbeta_{k}-\bbetabar\|_1\right\} : \\
&& \quad {\rm s.t.}\quad  \bz_k = \bA^{(k)}\bbeta_k  + \bc_{k} 
, k\in[K],
\end{eqnarray*}
where, for $\bz\in\R^{n_k}$, $G_k(\bz)=\sum_{i=1}^{n_k} f(z_{i})$ and $\bc_{k} = (c_1^{(k)},\ldots,c^{(k)}_{n_k})$ for any $k\in[K]$. \vskip5pt

\noindent We can now express the problem in min-max form as follows
\begin{eqnarray*}
\phi((\lambda_{1,k})_{k\in[K]},(\lambda_{2,k})_{k\in[K]})\!\!\!\! &=& \!\!\!\!\!\!\!\!
\min_{\substack{\bbetabar,\dbeta,\\\zz}}
\max_{\substack{\du,\dv,\\\dalpha}} \sum_{k=1}^K \bigg\{G_k(\bz_k) +
\balpha_k^T(\bA^{(k)}\bbeta_k +
 \bc_{k}  - \bz_k) \\
&& \qquad\qquad\qquad\qquad\qquad + \bu_k^T\bbeta_{k} +
\bv_k^T(\bbeta_{k}-{\bbetabar})\bigg\}\\
&& {\rm s.t.}\quad \|\bu_k\|_\infty\leq \lambda_{1,k},
\|\bv_k\|_\infty\leq\lambda_{2,k}, k\in[K].
\end{eqnarray*}
Now, solving for $\bbetabar,\dbeta$, we obtain
the dual constraints:
\begin{equation*}
\sum_{k=1}^K \bv_k={\bf 0}_{p}, \quad
{\bA^{(k)}}^T\balpha_k=\bu_k+\bv_k, \quad {\rm for}\ k\in [K].
\end{equation*}
Still denoting by $f^\star$ the Fenchel conjugate of the function
$f$, and eliminating variables $u_k$, $k=1,\ldots, K$, we obtain
the dual problem
\begin{eqnarray*}
\phi((\lambda_{1,k})_{k\in[K]},(\lambda_{2,k})_{k\in[K]})) =
\min_{\balpha_1,\ldots,\balpha_K} \sum_k \Big\{ \balpha_k^T\bc_k +
\sum_{i\in [n_k]} f^\star(\alpha_{k,i}) \Big\}
\end{eqnarray*}
under the constraints:
\begin{eqnarray*}
&\Big\|{\bA^{(k)}}^T\balpha_k - \bv_k \Big\|_\infty\leq \lambda_{1,k},\
\|\bv_k\|_\infty\leq \lambda_{2,k},\ \sum_k \bv_k ={\bf 0}_{p}\ {\rm and}\  \alpha_{k,i}\in {\cal D}{\rm om}(f^\star), \  {\rm for}\ k\in [K].
\end{eqnarray*}

For instance, in the logistic regression setting, this takes the
form
\begin{eqnarray*}
\phi((\lambda_{1,k})_{k\in[K]},(\lambda_{2,k})_{k\in[K]})) =
\min_{\balpha_1,\ldots,\balpha_K} \sum_{k\in[K]}
\balpha_{k}^T\log\balpha_k+(1-\balpha_k)^T\log(1-\balpha_k)
\end{eqnarray*}
under the constraints:
\begin{eqnarray*}
&\Big\|{\bA^{(k)}}^T\balpha_k - \bv_k \Big\|_\infty\leq \lambda_{1,k},\ 
\|\bv_k\|_\infty\leq \lambda_{2,k},\  \sum_k \bv_k ={\bf 0}_{p}, \ {\rm and}\  \balpha_{k}\in [0,1]^{n_k}, \  {\rm for}\ k\in[K].
\end{eqnarray*}
This problem is then equivalent to a standard entropy maximization and can therefore be solved using standard optimization toolboxes like the Mosek toolbox in R or Matlab for instance, which returns both optimal dual and primal solutions. (In the linear regression setting, the problem reduces to a quadratic programming and can also be solved with Mosek.)

\subsection{(Sparse) group lasso (spGroupLasso).}\label{sec:Group}
\cite{LouniciTsyb} and \cite{NegahbanWainwright} consider two different group lasso strategies. Their estimators are defined as minimizers of the following criterion
\begin{equation} 
 \sum_k \frac{\|\bY^{(k)} - \bX^{(k)}\bbeta^{(k)}\|_2^2 }{2n}+ \lambda \| \bB\|_{1,q}\label{eq:group}
\end{equation}
with typical values $q=2$ \citep{LouniciTsyb} or $q=\infty$ \citep{NegahbanWainwright}. The $\|\cdot\|_{1,q}$ encourages solutions $\widehat \bB$ to exhibit a row-wise group structure in the following sense: for each covariate $j\in[p]$, parameters $\hbeta^{(k)}_j$, for $k\in[K]$, are either all $0$ or all non-zero (the estimated effect of each covariate is either null for all tasks or non-null for all tasks).   

A slightly more flexible approach consists in adding an $L_1$ penalty to allow covariates to have a non-zero estimated effect on all but some tasks. This leads to the so-called sparse group lasso in standard regression models \citep{SimonSpGpLasso}. In our context, this suggests to minimize to following criterion
\begin{equation} 
\sum_k \frac{ \|\bY^{(k)} - \bX^{(k)}\bbeta^{(k)}\|_2^2 }{2 n}+ \lambda_1 \| \bB\|_{1,1} + \lambda_2 \| \bB\|_{1,q}.\label{eq:sparsegroup}
\end{equation}
Note that in Equation (\ref{eq:sparsegroup}) above solutions $\widehat \bB$ are encouraged to be sparse and to exhibit a group structure while in the Dirty model of \cite{Jalali} (see Equation (\ref{eq:dirty}) of the main paper) solutions are sums of two matrices: $\widehat{\bf S}$ that is encouraged to be sparse and $\widehat{\bf R}$ that is encouraged to have a group structure.

Non-asymptotic properties have been established for the non-sparse group-lasso in multi-task settings \citep{LouniciTsyb,NegahbanWainwright}. In particular, \cite{NegahbanWainwright} compare  the $L_1/L_\infty$-regularized method to IndepLasso in the special case ￼of $K=2$ linear regression problems with standard Gaussian designs whose supports have common size $s$ and overlap in a fraction $\alpha\in[0,1]$ of their entries. They focus on the capacity of each method to correctly specify the union of the two supports. Assuming a common number $n/2$ of observations for each task, they prove that  $L_1/L_\infty$-regularized method yields improved statistical efficiency if the overlap parameter is large enough ($\alpha\geq 2/3$), but has worse statistical efficiency than IndepLasso for moderate to small overlap ($\alpha< 2/3$). Recall that regarding sample complexity, the Dirty model was shown to strictly outperform both strategies under the same assumptions (except for extreme situations of full heterogeneity or full homogeneity where it matches the optimal strategy). Recall also that under the additional assumption $\beta_{1,j}^*\beta_{2,j}^*\geq 0$, we have established that our first approach $M_1$ performed at least similarly to the Dirty model (see Section \ref{sec:SampleComp} above).

\subsection{Additional results from the simulation study}\label{sec:Simulation_Supp}

\subsubsection{A simulation study when $K=2$}
\begin{figure}[h]
\begin{center}
\includegraphics[width=0.9\textwidth]{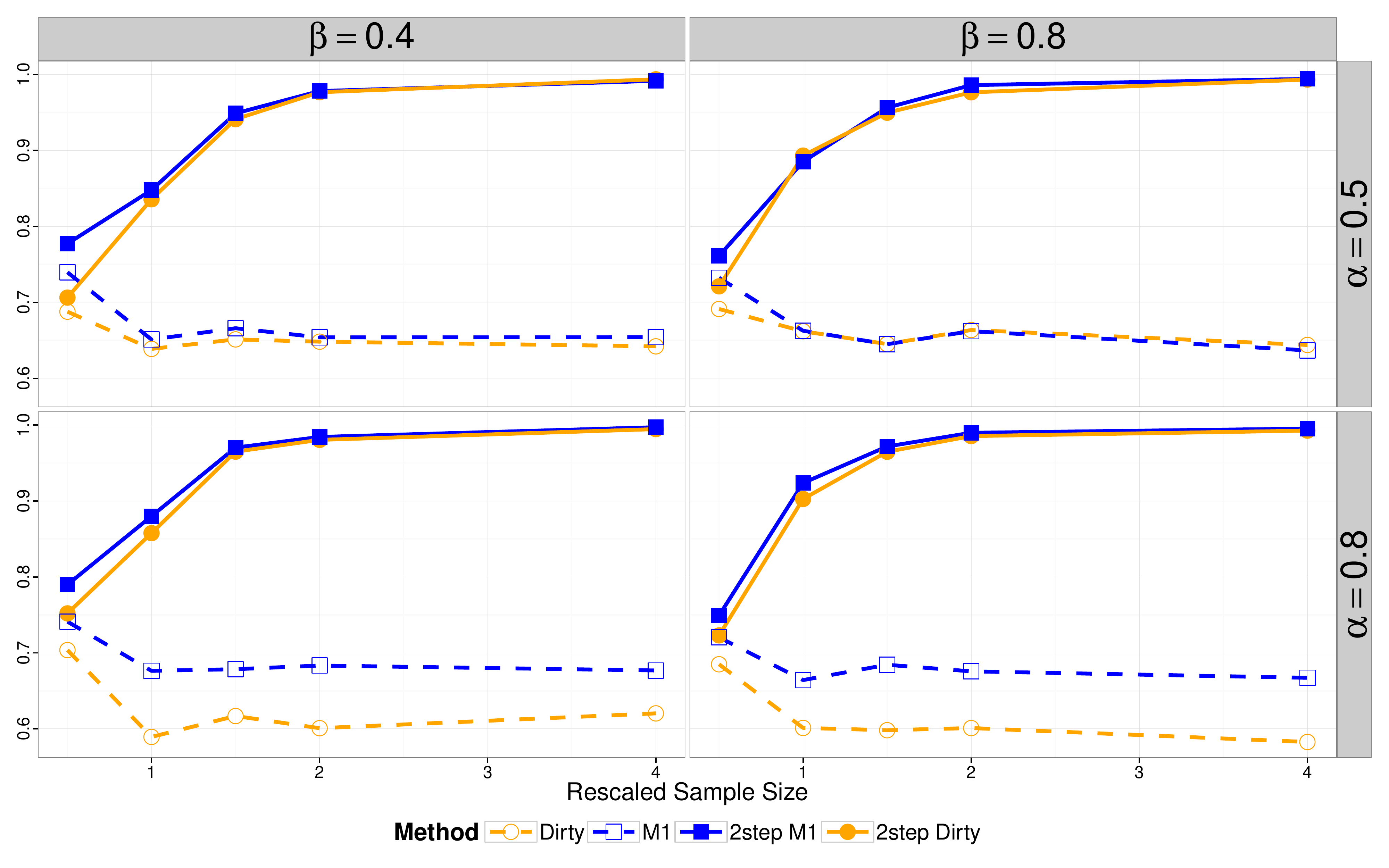}
\end{center}\caption{Results from the simulation study with $K=2$.}\label{fig:Lin2Tasks}
\end{figure}

Here we present a comparison of Dirty and $M_1$ when $K=2$. We actually mimic the simulation study presented in \cite{Jalali}. For a given value of $p$, the sparsity of both $\bbeta^*_1$ and $\bbeta^*_2$ is set to $s = \lfloor p/10\rfloor$. Results are presented for $p = 128$ (results for $p=256$ were similar). The proportion of non-zero and equal components in $\bbeta^*_1$ and $\bbeta^*_2$ is set to $\alpha\in\{0.5, 0.8\}$. For a given value of $\beta>0$, non-zero components are set to $\pm \beta$ with probability $1/2$. Design matrices were generated under a multivariate Gaussian ${\cal N}_p({\bf 0}_p, \bI_p)$ distribution, and noise variables under a ${\cal N}(0,0.1)$ distribution. Number of observations per stratum $n_1 = n_2 = n/2$ were set to $n_k = cs\log[p-(2\alpha)s]/2$, for $c\in\{0.5, 1, 1.5, 2, 4\}$ defining the rescaled sample size \citep{Jalali}. Selection of the optimal sparsity parameter was performed by using an independent test sample. Methods referred to as 2step Dirty and 2step $M_1$ further use a 2-step strategy to unshrunk estimates before computing the L2 prediction error on the test sample. Figure \ref{fig:Lin2Tasks} presents the results we obtained for the support recovery accuracy. It confirms that Dirty and $M_1$ perform similarly.

\subsubsection{Support recovery accuracy for the methods compared on Figure \ref{fig:L2Pred.simToep} of the main paper}
\begin{figure}[h]
\begin{center}
\includegraphics[width=0.9\textwidth]{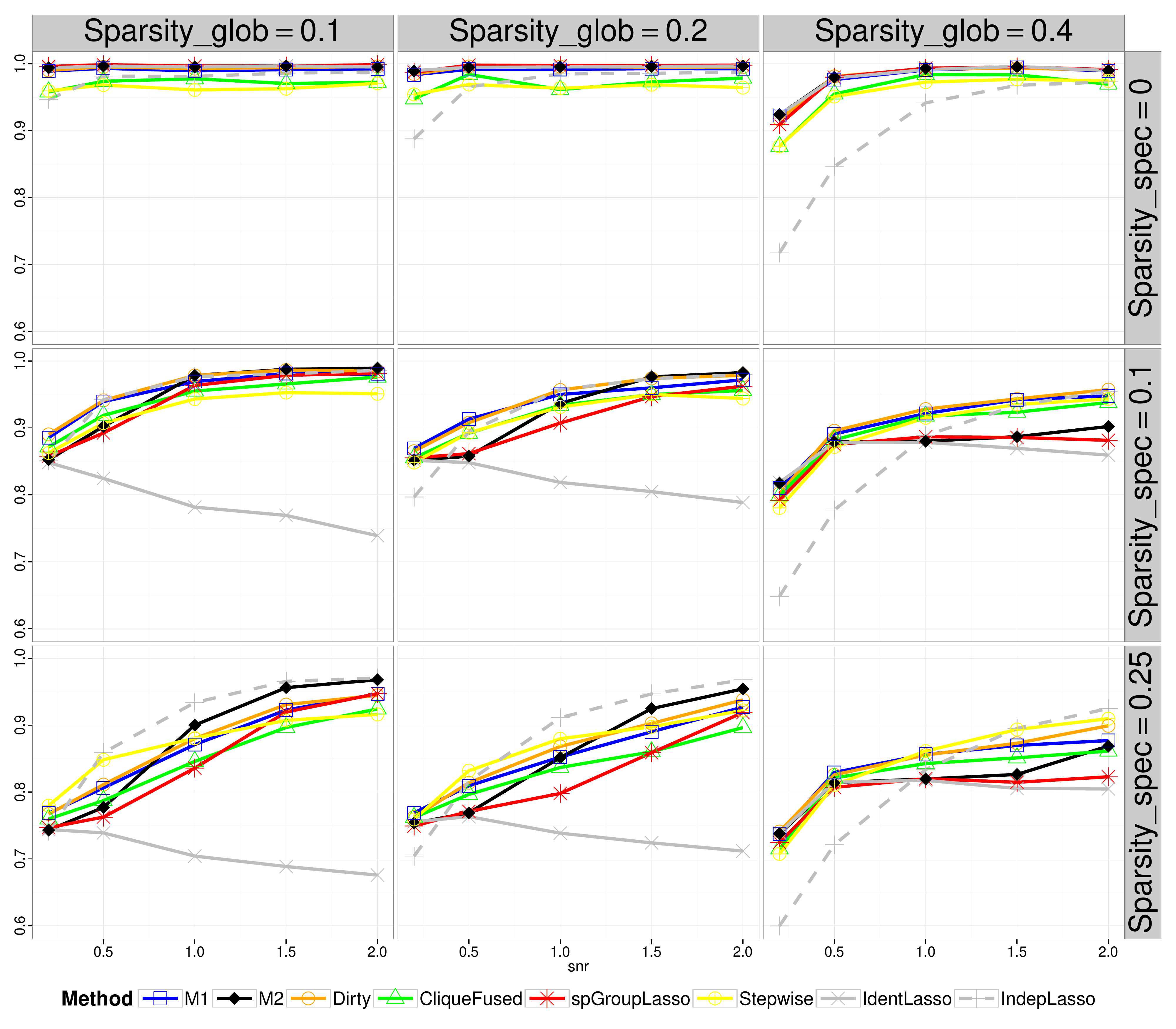}
\end{center}\caption{Results from the first simulation study: support recovery accuracy.}\label{fig:Acc.simToep}
\end{figure}

Figure \ref{fig:Acc.simToep} presents the support recovery accuracies for the methods compared in terms of prediction accuracy on Figure \ref{fig:L2Pred.simToep} of the main paper. As mentioned in the main paper, conclusions drawn from Figure \ref{fig:Acc.simToep} are very similar to those drawn from Figure \ref{fig:L2Pred.simToep}  of the main paper. 

\subsubsection{Performance of $M_2$ depending on the $(\lambda_1,\lambda_2)$ grid}\label{sec:LambdaGrid}
All methods presented in the main article involve tuning parameters that need to be carefully selected in practice. Generally speaking, a predefined grid of $\lambda$ values has first to be constructed. For instance in the single stratum setting, the minimal value $\lambda_{\rm max}$ ensuring that the vector returned by the lasso is null for all $\lambda\geq \lambda_{\rm max}$ has a closed-form expression ({\em e.g.}, for linear and logistic regression) and the grid is then generally of the form $\{\lambda_{\rm max}/1000, \ldots, \lambda_{\rm max}\}$. Most methods considered in the main paper involve two tuning parameters, and the derivation of $\lambda_{1,{\rm max}}$ and $\lambda_{2,{\rm max}}$ is not straightforward. Because our simulated data are such that $X^{(k)}_j \sim {\cal N}(0,1)$ for all $k\in[p]$ and $j\in[p]$, and $n_k = n/K$ for all $k\in[K]$, we limit the presentation to the case where tuning parameters do not depend on $k$; if the $n_k$'s are not equal, one solution for $M_1$ consists in first standardizing the columns of matrix $\bcX$ and then proceed as described here. For the Dirty approach of \cite{Jalali} we proceed as they did in their simulation study. Because of its connection with the lasso, the following strategy can be used for our first approach $M_1$ (or its adaptive version). For any pair $(\lambda_1,\lambda_2)$ such that $\lambda_1\geq K\lambda_2$, it is easy to see from Equation (\ref{eq:RewritingM1}) of the main paper that the estimated common effect $\hbbetabar$ will be ${\bf 0}_p$ and $M_1$ then reduces to a version of IndepLasso ran with the tuning parameter set to $\lambda_2$. On the other hand, if $\lambda_1=0$ then $\bbetabar$ is unpenalized, and $M_1$ returns parameter vectors $\hbbeta_k = \hbbetabar$ for $\lambda_2$ large enough. Therefore, our strategy is to first make the $\lambda_1/\lambda_2$ ratio vary on the interval $[0,K]$ (we take 50 equally-spaced values on a log-scale), and for each value of this ratio $r$, the glmnet function can be used to compute the $\lambda_{2,{\rm max}}(r)$ value, along with the 50 models returned with $\lambda_2(r)$ varying on the grid of 50 equally-spaced values (on a log-scale) $\{\lambda_{2,{\rm max}}(r)/1000,\ldots,\lambda_{2,{\rm max}}(r)\}$. 

\begin{figure}[h]
\begin{center}
\includegraphics[width=0.9\textwidth]{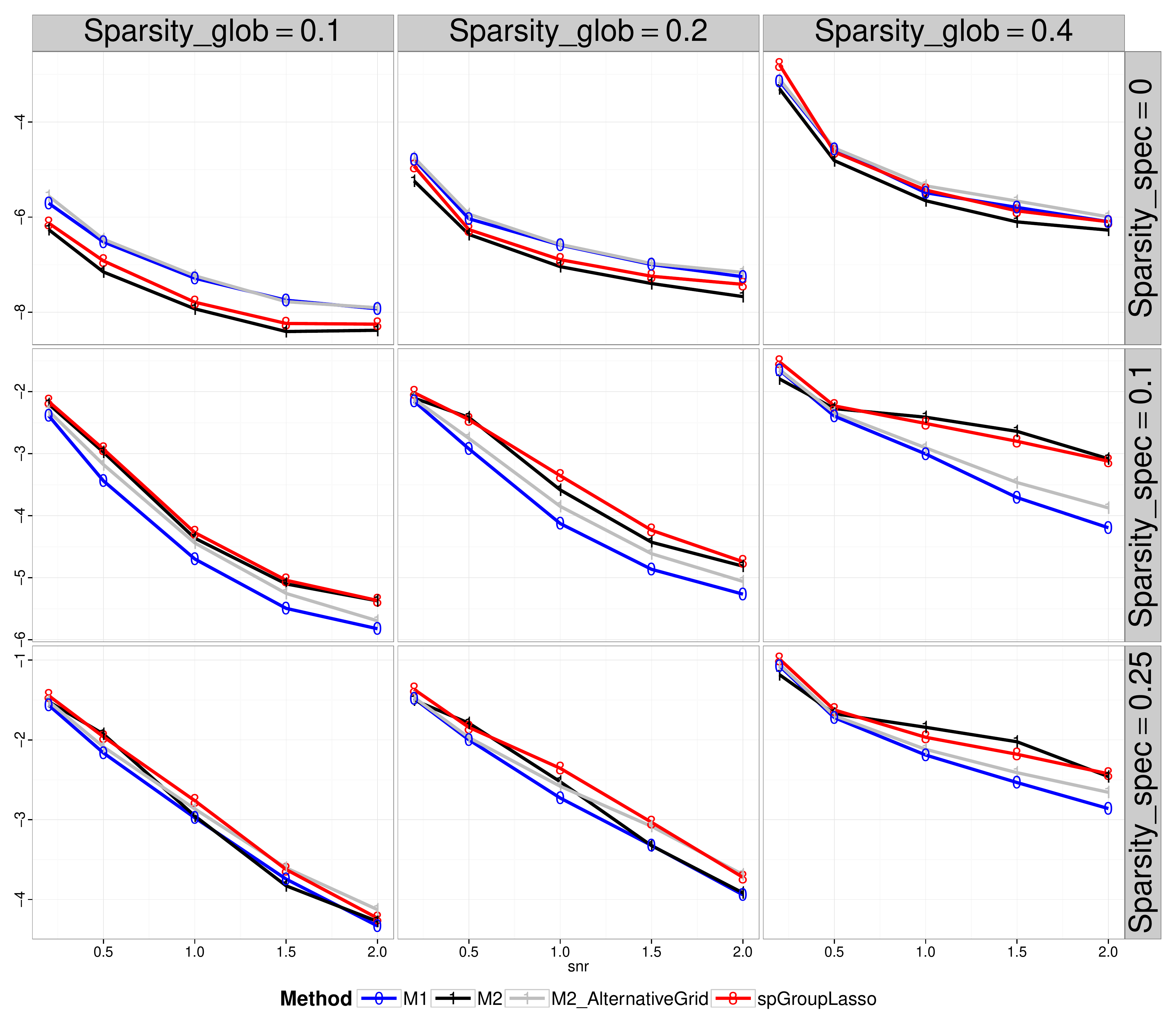}
\end{center}\caption{Comparison of the prediction performance for $M_2$ depending on the strategy used to construct the $(\lambda_1, \lambda_2)$ grid of values. }\label{fig:CompGrid}
\end{figure}

For the other methods ($M_2$, CliqueFused and spGroupLasso) we proceed as follows. We first consider the case $\lambda_2 = 0$. The methods then all reduce to a standard lasso for which the $\lambda_{1,{\rm max}}$ value can be computed. The grid of $\lambda_1$-values is then chosen as the set of 50  values $\{\lambda_{1,\rm max}/1000, \ldots, \lambda_{1,\rm max}\}$, equally-spaced on a $\log$-scale. Then, for each $\lambda_1$ value on this grid, the minimal value $\lambda_{2,{\rm max}}(\lambda_1)$ is numerically approximated. This value is such that for all $\lambda_2 \geq \lambda_{2,{\rm max}}(\lambda_1)$, the considered method computed with parameters $(\lambda_1,\lambda_2)$ returns a vector of parameter $\hbbeta(\lambda_1,\lambda_2)$ with the lowest possible overall complexity. In particular, for methods $M_1$, $M_2$ and CliqueFused, this means that $\hbbeta_k(\lambda_1,\lambda_2) = \hbbeta_{k'}(\lambda_1,\lambda_2)$ for all $(k,k')\in[K]$ and $\lambda_2 \geq \lambda_{2,{\rm max}}(\lambda_1)$. Such an approximate $\lambda_{2,{\rm max}}(\lambda_1)$ value is simply obtained by iteratively running the considered method for various values of $\lambda_2$, starting from a huge value, and successively dividing it by 2 for instance. 
Another, more simple, strategy is as follows. We first consider the case $\lambda_2 = 0$. The methods then all reduce to a standard lasso for which the $\lambda_{1,{\rm max}}$ value can be computed. Then, we use the grid of 50 values $\{\lambda_{1,\rm max}/1000, \ldots, \lambda_{1,\rm max}\}$ for both $\lambda_1$ and $\lambda_2$. This strategy is referred to as AlternativeGrid on Figure  \ref{fig:CompGrid} below which presents the comparison of the performance for $M_2$ implemented with the two strategies on the first simulation study. The comparison of the black and grey curves illustrates how difficult it is to empirically compare methods. When using the AlternativeGrid strategy, $M_2$ performs similarly to $M_1$, while using the other strategy, it performs like SpGroupLasso (which uses the same strategy for the grid construction). Therefore, the observed difference of performance between the various strategies can be (at least partly) explained by the choice for the initial grid of $(\lambda_1, \lambda_2)$ values.

\end{document}